\documentclass[reprint,amsmath,amssymb,aps,prb,nofootinbib,longbibliography]{revtex4-2}
\usepackage{amsmath,amssymb}
\usepackage[dvipsnames]{xcolor}
\usepackage{hyperref}
\usepackage[capitalise]{cleveref}
\usepackage{mathtools}
\usepackage{siunitx}
\usepackage[caption=false]{subfig}
\crefrangeformat{equation}{Eqs.~(#3#1#4-#5#2#6)}
\crefmultiformat{equation}{Eqs.~(#2#1#3}%
{,~#2#1#3)}{,~#2#1#3}{,~#2#1#3)}

\begin{document}
\preprint{APS/123-QED}

\title{The line bundle regime and the scale-dependence \\of continuum dislocation dynamics}

\author{Joseph Pierre Anderson}\altaffiliation[Currently at: ]{Department of physics and astronomy, Carthage College.}\email{janderson19@carthage.edu}
 \affiliation{School of materials engineering, Purdue University}
\author{Anter El-Azab}
\affiliation{School of materials engineering, Purdue University}%

\date{\today}

\begin{abstract}

Continuum dislocation dynamics (CDD) has become the state-of-the-art theoretical approach for mesoscale dislocation plasticity of metals. Within this approach, there are multiple CDD theories that can all be derived from the principles of statistical mechanics. In these theories density-based measures are used to represent dislocation lines. Establishing these density measures requires some level of coarse graining with the result of losing track of some parts of the dislocation population due to cancellation in the tangent vectors of unaligned dislocations. The leading CDD theories either treat dislocations as nearly parallel, as in the line bundle approach [J. P. Anderson and A. El-Azab, Phys. Rev. B \textbf{109} (2024)], or distributed locally over orientation space [Hochrainer, Phil. Mag. \textbf{95}, 1321 (2015)]. The difference between these theories is a matter of the spatial resolution at which the definition of the relevant dislocation density field holds: for fine resolutions, single dislocations are resolved and there is no cancellation; for coarse resolutions, whole dislocation loops could contribute at a single point and there is complete cancellation. In the current work, a formulation of the resolution-dependent transition between these limits is presented in terms of the statistics of dislocation line orientation fluctuations about a local average line direction. From this formulation, a study of the orientation fluctuation behavior in intermediate resolution regimes is conducted. The fluctuation distributions are found to be roughly Cauchy in shape and broaden with increased coarse-graining length. Two possible closure equations for truncating the moment sequence of the fluctuation distributions relating the two theories mentioned above are evaluated from data, the newly introduced line bundle closure and the previous standard maximum entropy closure relations. The line bundle closure relation is shown to be accurate for coarse-graining lengths up to half the dislocation spacing and the maximum entropy closure is found to poorly agree with the data at all coarse-graining lengths.

\end{abstract}

\maketitle

\section{Introduction}
The problem of strain hardening in metals is one of the last frontiers of classical physics. A first-principles solution to this problem would involve a dynamical theory of dislocations---the line defects that carry plastic deformation---capable of predicting hardening coefficients present in phenomenological relations for the flow stress \cite{kocksPhysicsPhenomenologyStrain2003}. These hardening coefficients in turn seem to be strongly tied to the dynamic self-organization of the dislocations into various spatial patterns \cite{sauzayScalingLawsDislocation2011}. Continuum dislocation dynamics, which models the transport of various dislocation density measures representing  the local dislocation state of the crystal, seems to be a promising way forward toward a first-principles theory of dislocation patterning and, eventually, strain hardening. However, the so-called statistical storage problem—the tendency for any continuum dislocation density measure to underdetermine the local plastic deformation—has been a perennial impediment to the development of these theories, a statement as true today as it was when Kr\"{o}ner pointed it out in his final outlook on the field \cite{Kroner2001}. 

Upon beginning to set to the problem of continuum dislocation dynamics, one is immediately faced with a choice: which density measures to use? Different approaches to statistical storage motivate different answers to this question and naturally lead to substantially different theories of continuum dislocation motion.
This pernicious difference makes it difficult for researchers, especially the practitioners of the respective theories, to recognize the worthwhile and complementary contributions these theories make to the physics of plasticity or how the theories are related. The current work is intended as a bridge between two such frameworks of continuum dislocation dynamics, the line bundle  \cite{andersonDislocationCorrelationsContinuum2024} and the higher order  \cite{Hochrainer2015} formulations. Hence, it is worth describing the statistical storage problem in more detail and how these formulations address it differently.

Continuum dislocation fields, by their nature, treat a collection of dislocations in a representative volume around a single field point. Each of these dislocations has a line direction $\boldsymbol{l}$ and a Burgers vector $\boldsymbol{b}$. Taken together, these are related to the incompatibility of a dislocation which gives rise to long-range mechanical fields. The incompatibility (or Kr\"{o}ner-Nye) tensor field is simply the net incompatibility of all dislocations in the local collection:
$$
\boldsymbol{\alpha}(\boldsymbol{r}) = \sum \boldsymbol{l}\otimes \boldsymbol{b}.$$
Its relationship to mechanics led some to develop dynamical theories of the evolution of the incompatibility tensor \cite{muraContinuousDistributionMoving1963}, and indeed some are still pursuing such theories \cite{Acharya2006,Arora2020}. However, one of the difficulties identified by Kr\"{o}ner \cite{Kroner2001} is simply that in the summation over any collection of dislocations, cancellation can occur in either of these vector quantities. The incompatibility tensor thus does not describe all of the dislocation content of a collection, only its so-called geometrically necessary dislocation (GND) content, which, in Nye's sense \cite{Nye1953}, represents the net dislocation content needed to describe the mismatch of crystalline orientations across this volume. The dislocations are the dynamic objects, and so incompatibility has to supplement the motion of the GND content with a phenomenological term guessing the evolution of the unaccounted dislocations \cite{Acharya2006}. 

The fact that dislocations belong to well-defined crystallographic slip systems (combinations of Burgers vectors and slip planes to which the line direction is generally confined) suggests that there are two forms of cancellation in the incompatibility tensor that should be distinguished. If the dislocation collection is made up of enumerated ‘species’ of dislocations belonging to distinct slip systems, cancellation might occur when we a) obtain the average line direction on a single slip system (this subcollection shares a Burgers vector), or b) when summing over the various slip systems to obtain the net incompatibility. The latter is a delightfully geometric crystallography problem amounting to exploring the nullspace of a projection matrix and is given a definitive treatment in face-centered cubic (fcc) crystals by Arsenlis and Parks \cite{Arsenlis1999}. However, this issue can be simply sidestepped by considering a separate dislocation density field pertaining to each slip system \cite{El-Azab2018}. The former issue, the cancellation of line directions within a slip system population, still remains a nuanced problem to say the least.

The single-slip statistical storage problem essentially boils down to the question of when it is relevant. This problem can be traced back to two-dimensional theories of straight edge dislocations in the investigation of x-ray diffraction line profile analysis \cite{wilkensTheoreticalAspectsKinematical1970,Groma1998}. Biting the bullet of the single-slip problem involves considering dislocation subcollections of different direction with separate density fields; in two dimensions, this simply involves partitioning the dislocation population into positive and negative dislocations \cite{Groma1998,Groma1999,Zaiser2001,Groma2003}. The fact that such a toy model was already a rich playground for the investigation of collective dislocation phenomena \cite{ispanovityEvolutionCorrelationFunctions2008,dusanispanovityCriticalityRelaxationDislocation2011,dusanispanovityAvalanches2DDislocation2014,Ispanovity2017,Wu2018,wuCellStructureFormation2018,wuCyclicloadingMicrostructurepropertyRelations2019,Ispanovity2020} is evidence of the subtlety of the single-slip statistical storage problem. However, in three dimensions, the space of dislocation orientations is no longer trivial: the distribution over positive and negative orientation must be replaced with a distribution over a continuous orientation space (the unit circle for planar dislocations as in the case of fcc crystals) \cite{El-Azab2000a}. It was soon realized that this further complicates matters by also needing to include a density measure related to the curvature of dislocations of a given orientation in the collection \cite{Hochrainer2007}. This high-dimensional theory was subsequently simplified by considering only low-order moments of the orientation distribution \cite{Hochrainer2015}, and a commensurate analytical treatment of the energetics also followed shortly after \cite{Zaiser2015}. While this is a powerful theoretical framework, it is complex. This complexity leads to difficulties in closing the theory simply due to the difficulty of comparing the theory to discrete dislocation data or even underlying dislocation processes \cite{SandfeldPo2015,Monavari2016,Monavari2018,Sandfeld2019,Sudmanns2019,Sudmanns2020,Song2021,Hochrainer2022}. Due to this difficulty, the current trajectory of this theory is towards large length scale, phenomenological crystal plasticity especially in relationship to phase field plasticity theories \cite{Groma2015,Groma2021,zhangContinuumDislocationDynamics2025}.

One might question if all this theoretical machinery is necessary, or equivalently, whether single-slip statistical storage is always relevant. If it is not, the orientation distribution is single-valued at all points in space (all dislocations are parallel), analogously to a smooth bundle of lines. This possibility is primarily motivated by length scale considerations; if the density is resolved on a small scale,  often taken to be ~10-100 nm, there is an elastic alignment mechanism as well as an annihilation mechanism for oppositely signed dislocations. The assumption that the vector density accurately captures the line orientation of the local dislocation collection has several benefits. First, it admits a more straightforward dynamical analogy with discrete dislocations (as the streamlines of the density field \cite{Lin2021a}), e.g., for considerations of dislocation reactions \cite{Weger2019,Lin2020,Vivekanandan2021,vivekanandanDataDrivenApproach2023} and average driving forces \cite{Anderson2021,andersonDislocationCorrelationsContinuum2024}. Moreover, because of the detailed orientation description and its close tie to the incompatibility tensor, the mechanics of finite deformation—--which are necessary to describe the lattice rotations central to theories of strain hardening \cite{kocksPhysicsPhenomenologyStrain2003,sauzayScalingLawsDislocation2011}—--and their influence on the dislocation dynamics are more suitably described using the vector density theory \cite{Starkey2020,Hochrainer2020}. Due to the fine spatial resolution of the theory, the line bundle framework is sometimes maligned as a ``pseudo-continuum theory'' with no benefit surpassing discrete dislocation dynamics, which operates at a similar length scale \cite{Monavari2018,Hochrainer2022,Groma2021}. However, the stable computational complexity \cite{Lin2021} and formulation of finite deformation mechanics \cite{Starkey2020,starkeyTotalLagrangeImplementation2022} allow it to push to higher strain regimes than discrete dislocation dynamics. Moreover, the theory is a rich dynamical system in its own right, showing promise of studying the onset of dislocation patterns \cite{Xia2015}. 

So how should these approaches be reconciled? It is well-known that the vector density dynamics are recoverable as a special case of the high-dimensional dynamics \cite{Sedlacek2010,Hochrainer2010}. If the question is one of length scales, where is the transition between the vector density theory (which is obviously correct in the limit of finer scales) and the higher-order theory (which is obviously necessary in the limit of coarser scales)? Because both theories make claims regarding the orientation distribution, this question should, in principle, be answerable by studying the orientation statistics of dislocations. The present work sets out to accomplish this comparison by defining local orientation distribution functions in a spatial averaging framework that makes length scale explicit (section \ref{sec:2_localcollections}), situating points of comparison between the line bundle and higher-order theories in a theoretical analysis of orientation fluctuations (\ref{sec:3_reduced_descriptions}), and then evaluating this relationship between the line bundle and higher-order theories from statistical analysis of orientation fluctuations in discrete dislocation configurations (\ref{sec:4_calculation}-\ref{sec:5_results}). Implications of the present work for the field of continuum dislocation dynamics are discussed in section \ref{sec:6_disc}.

\section{A full description of local dislocation collections \label{sec:2_localcollections}}
In this section, we present a higher-order continuum theory of
dislocations along the lines of~\cite{Hochrainer2007}. This framework considers a collection of dislocations at a point in space by means of a
distribution over an orientation space, taken to be the unit circle in
the case of pure glide systems. Although the theory itself is agnostic
to the averaging process used to define a local collection of
dislocations, we will present this theory in the context of a spatial
averaging of dislocations (as opposed to ensemble averaging). We make
this choice because it elucidates the scale dependence of the theory
that will be the focus of the present work.

When attempting to model the dislocation state of a crystal by means of
a spatially varying dislocation density field, there is always an
implicitly associated length scale. This implicit length scale is the
characteristic length scale of gradients in the field. Even at the level
of `discrete' dislocations, it is often instructive to think of
dislocations as a density field. The core region is properly treated by means of a Burgers vector distribution function which smears the singular dislocation by convolution of the discrete line with some kernel
function determined by the lattice itself~\cite{Po2018,Cai2006}. Such a smearing occurs on distances on the order of a few lattice constants (5-10 \AA). This treatment of the discrete
dislocation line determines what we might consider the `true' strength of
gradients in the physical dislocation density field. Any dislocation density
field which varies on distances \(L\) greater than this fundamental core
distance can then be thought of as an additional smearing of the
dislocation with kernels (\(w_{L}\)) of unit norm
(\(\int w_{L}\left( \boldsymbol{r} \right)d^{3}\boldsymbol{r }=1\)) and range
\(L.\) This means that the dislocation density field no longer describes
\emph{discrete} dislocations but rather a \emph{collection} of
dislocations which are contained in a characteristic volume of
approximate scale \(L\).

In the above interpretation of a density field, we can discuss several
characteristics of this collection of dislocations which are implicitly
described by the density field at each point. To elucidate this fact, let
us introduce the following notation. The set of
dislocations\(\mathcal{\ L}\) is a directed line manifold embedded
within a crystal manifold \(\mathcal{M}\), which for our present
purposes we take to be equivalent to \(\mathbb{R}^{3}.\ \)Consider at
present only glide dislocations on a single slip system. For simplicity,
consider the smearing kernel \(w_{L}(\boldsymbol{r -}\boldsymbol{r}_{0}\)) to be
of compact support \(\Omega_{\boldsymbol{r}_{0}}^{(L)}\mathcal{\subset M}\).
Suppose that there is some subset of the dislocation state which
intersects \(\Omega_{\boldsymbol{r}_{0}}^{(L)}\):
\begin{equation}\Lambda_{\boldsymbol{r}_{0}}\coloneqq\mathcal{L}\cap\Omega_{\boldsymbol{r}_{0}}^{(L)}\end{equation}
This is the collection of dislocations described by the density field at
\(\boldsymbol{r}_{0}\). The smearing kernel, or weight function, allows us
to define two important concepts, namely the scalar dislocation density
at a point:
\begin{equation}\rho( \boldsymbol{r}_{0} )\coloneqq\int_{\Lambda_{\boldsymbol{r}_{0}}^{(L)}}^{}{w_{L}\left( \boldsymbol{r}_{l}-\boldsymbol{r}_{0} \right)dl},\end{equation}
and local averages of the dislocation system:
\begin{equation}
\left\langle A(\boldsymbol{r}_l)\right\rangle \coloneqq \frac{1}{\rho(\boldsymbol{r}_0)}\int_{\Lambda_{\boldsymbol{r}_{0}}^{(L)}}^{}{w_{L}\left( \boldsymbol{r}_{l}-\boldsymbol{r}_{0} \right) A(\boldsymbol{r}_l) dl},\end{equation}
where \(A\left( \boldsymbol{r}_{l} \right)\) is some local property of a dislocation line. That is, we can treat the weight function once normalized by the total dislocation density as a probability density on the dislocation line.

\subsection{Defining the orientation distribution
function}

To begin defining the sort of local properties we will be interested in, let us define the orientation of a dislocation line. We will consider the direction of these planar dislocation lines (a consequence of the glide assumption) in several ways. First, we can consider the line orientation as a vector by considering the derivative of the map by which the dislocation line is parametrically defined:
\begin{subequations}
\begin{align}
    \mathcal{L} &\colon\ \mathbb{R}\to\mathcal{M}, & \boldsymbol{r}_l(s) \in \mathcal{L},\\
    \boldsymbol{l}(\boldsymbol{r}_l)&\coloneqq \frac{\partial_s\boldsymbol{r}_l(s)}{\vert\partial_s\boldsymbol{r}_l(s)\vert}.
\end{align}
\end{subequations}

Since line orientation will be of central importance to the remainder of this article, we will also introduce the following interchangeable representations of the orientation. Because this is a planar set of dislocations on a known slip plane with normal \(\hat{\boldsymbol{n}}\) with a fixed Burgers vector  \(b\hat{\boldsymbol{b}}\) lying in that plane, there is a natural coordinate system for the dislocation unit tangent vector. We can define this coordinate representation by a single angular parameter. This angle is defined counter-clockwise around the normal direction with zero at the Burgers vector direction. In this coordinate form, the unit tangent is given by:
\begin{subequations}
    \begin{align}
    \boldsymbol{l}\left( \varphi\left( \boldsymbol{r}_{l} \right) \right)&=\cos{\varphi\left( \boldsymbol{r}_{l} \right)}\hat{\boldsymbol{b}} + \sin{\varphi\left( \boldsymbol{r}_{l} \right)}\hat{\boldsymbol{a}},\label{eq:orientationDef}\\
    \hat{\boldsymbol{a}}&\coloneqq\hat{\boldsymbol{n}}\times\hat{\boldsymbol{b}}.
\end{align}
\end{subequations}
where \(\hat{\boldsymbol{a}}\) would be the line direction of a positive pure edge dislocation. Due to the nature of orientation averages, it is convenient to deal with the unit complex number corresponding to this angle. In this case, the coordinate representation of the unit tangent is given by
\begin{subequations}\label{eq:orientationExponential}
    \begin{align}
\boldsymbol{l}\left( z\left( \boldsymbol{r}_{l} \right) \right)&=\Re\left[ z\left( \boldsymbol{r}_{l} \right)\left( \hat{\boldsymbol{b}} - i\hat{\boldsymbol{a}} \right) \right] \label{eq:zdir}\\
z\left( \boldsymbol{r}_{l} \right) &\coloneqq e^{i\varphi\left( \boldsymbol{r}_{l} \right)}.
\end{align}
\end{subequations}
with $\Re$ referring to the real part of $\left[\cdot \right] $. Now that we have a better understanding of the orientation parameter, we can consider the likelihood that a dislocation segment in \(\Lambda_{\boldsymbol{r}_{0}}^{(L)}\) has an orientation falling in some subset $B$ of the unit circle \(S^{1}\). This is inherited from our spatial average in Eq. (1) by:
\begin{equation}P_{\boldsymbol{r}_{0}}^{(L)}\left( \varphi \in  B  \right) \coloneqq \frac{\int_{\Lambda_{\boldsymbol{r}_{0}}^{(L)}}^{}{1_{B}\left( \varphi\left( \boldsymbol{r}_{l} \right) \right)w_{L}\left( \boldsymbol{r}_{l} - \boldsymbol{r}_{0} \right)dl}}{\rho\left( \boldsymbol{r}_{0} \right)}, \end{equation}
where $1_{B}$ is the indicator function on the set $B$. Under quite general conditions, this probability measure can be
represented by a density function which we will call the orientation
distribution function $g(\varphi)$:

\begin{equation}
P_{\boldsymbol{r}_{0}}^{(L)}(\varphi \in B) = \int_{B}^{}{g_{\boldsymbol{r}_{0}}^{(L)}(\varphi)d\varphi}.\label{eq:2odfDef}
\end{equation}

Now the local average (in a neighborhood of size \(L\) around the point \(\boldsymbol{r}_{0}\)) of any orientation dependent line property can be represented by this orientation distribution function instead of the spatial distribution:
\begin{equation}\left\langle A\left( \varphi\left( \boldsymbol{r}_{l} \right) \right) \right\rangle_{\boldsymbol{r}_{0}}^{(L)} = \int_{S^{1}}^{}{A(\varphi)g_{\boldsymbol{r}_{0}}^{(L)}(\varphi)d\varphi}.\end{equation}
This orientation distribution function suffers one weakness: the orientation parameter \(\varphi\) is tied to a global coordinate system and is thus not well-suited to comparing local collections of dislocations at different points. We can remedy this by defining a canonical local coordinate system based on the first circular average (which can be performed in terms of \(\boldsymbol{l}(\varphi)\) or \(z(\varphi)\) but not directly in terms of \(\varphi\)):
\begin{equation}
{\overline{z}}_{\boldsymbol{r}_{0}}^{(L)} = \left\langle z(\varphi) \right\rangle_{\boldsymbol{r}_{0}}^{(L)}.\label{eq:3zav}
\end{equation}
Note that, in general, the magnitude of this average is less than or equal to unity.\footnote{This is responsible for the single-slip statistical storage of dislocation content.} We will refer to this magnitude as the polarization
\(\beta_{1}\):
\begin{equation}\beta_{1}^{(L)}\left( \boldsymbol{r}_{0} \right) \coloneqq \left| \left\langle z(\varphi) \right\rangle_{\boldsymbol{r}_{0}}^{(L)} \right| \leq 1\end{equation}
as it represents the fraction of the total dislocation content which is
not cancelled by the vector sum of dislocations in the volume   $\Omega_{\boldsymbol{r}_{0}}^{(L)}$. That is, the so-called `geometrically necessary' dislocation content is given by:
\begin{equation}\rho_\text{GND}\left( \boldsymbol{r} \right)=\beta_{1}\left( \boldsymbol{r} \right)\rho\left( \boldsymbol{r} \right)\end{equation}
The average orientation, on the other hand, which will define our local coordinate system, is defined by the renormalized circular average so as to retain its quality as a unit vector. This can be defined in equivalent ways:
\begin{subequations}
    \label{eq:4avorient}
    \begin{align}    {\overline{\boldsymbol{l}}}_{\boldsymbol{r}_{0}}^{(L)} &\coloneqq {\frac{1}{\beta_{1}}\left\langle \boldsymbol{l}(\varphi) \right\rangle}_{\boldsymbol{r}_{0}}^{(L)} \\
    &\coloneqq\frac{1}{\beta_{1}}\Re\left\lbrack {\overline{z}}_{\boldsymbol{r}_{0}}^{(L)}\left( \hat{\boldsymbol{b}}\boldsymbol{-}i\hat{\boldsymbol{a}} \right) \right\rbrack.
\end{align}
\end{subequations}
Similarly, the orthogonal direction in the slip plane can be defined:
\begin{subequations}
    \begin{align}
    {\overline{\boldsymbol{p}}}_{\boldsymbol{r}_{0}}^{(L)} &\coloneqq \hat{\boldsymbol{n}} \times {\overline{\boldsymbol{l}}}_{\boldsymbol{r}_{0}}^{(L)} \\
    &=-\frac{1}{\beta_{1}}\Im\left\lbrack {\overline{z}}_{\boldsymbol{r}_{0}}^{(L)}\left( \hat{\boldsymbol{b}}\boldsymbol{-}i\hat{\boldsymbol{a}} \right) \right\rbrack\\
    &=\frac{1}{\beta_{1}}\left\langle \boldsymbol{l}\left( \varphi + \pi/2 \right) \right\rangle_{\boldsymbol{r}_{0}}^{(L)}.
\end{align}
\end{subequations}
Thus, we have a local orthonormal coordinate system \(\overline{\boldsymbol{l}},\overline{\boldsymbol{p}},\hat{\boldsymbol{n}}\) defined in terms of the local dislocation arrangement. Either unit vector can then define an average angle:
\begin{subequations}
    \begin{align}
    \boldsymbol{l}\left( {\overline{\varphi}}_{\boldsymbol{r}_{0}}^{(L)} \right) &\coloneqq {\overline{\boldsymbol{l}}}_{\boldsymbol{r}_{0}}^{(L)},\label{eq:averageangleDef}\\
    \boldsymbol{l}\left( {\overline{\varphi}}_{\boldsymbol{r}_{0}}^{(L)} + \pi/2 \right) &\coloneqq {\overline{\boldsymbol{p}}}_{\boldsymbol{r}_{0}}^{(L)}.
\end{align}
\end{subequations} 

We briefly digress at this point to mention that the first circular average (and therefore our local coordinate system) is a central quantity in defining the internal elastic fields caused by dislocations. The classical continuum theory of dislocations~\cite{DeWit1973} is based on the additive decomposition of the displacement gradient\footnote{Properly speaking, there should be a multiplicative decomposition of the total deformation gradient. This additive decomposition is only appropriate for infinitesimal strains. Cf. \cite{starkeyTotalLagrangeImplementation2022} for a treatment of finite strains using the multiplicative decomposition.} into an elastic and plastic distortion fields, distinguished by the fact that the symmetric portion of the elastic distortion gives rise to stresses in the material:
\begin{align}
\nabla\boldsymbol{u}&=\mathbf{D}^{(e)}+\mathbf{D}^{(p)}\\
\boldsymbol{\sigma} &=\mathbb{C} : \frac{1}{2}\left(\mathbf{D}^{(e)} + \mathbf{D}^{(e)T}\right)\\
&=\mathbb{C} : \mathbf{D}^{(e)}
\end{align}
since the stiffness tensor \(\mathbb{C}\) is symmetric. The curl of each distortion field is given by the incompatibility density (Kr\"{o}ner-Nye) tensor \cite{kroner1981continuum}:
\begin{equation}\nabla\times\mathbf{D}^{(e)}= - \nabla \times\mathbf{D}^{(p)}= \boldsymbol{\alpha}\end{equation}
and the incompatibility density tensor itself is determined by the dislocation arrangement (in the case of a single slip system) as:
\begin{equation}\boldsymbol{\alpha}\left( \boldsymbol{r} \right)\coloneqq\rho\left( \boldsymbol{r} \right)\beta_{1}\left( \boldsymbol{r} \right){\overline{\boldsymbol{l}}}_{r}\boldsymbol{\otimes}b\hat{\boldsymbol{b}}\end{equation}
Thus, we can see that even in a basic solution to the mechanics, the orientation distribution function has a central place, although only in terms of the first circular average discussed above.

Before moving on to how the orientation distribution function affects the collective dynamics of dislocations, we pause to note that we will, for the remainder of the present work, mostly consider the `mean-centered' distribution (in a circular sense). This distribution, which we will term the orientation fluctuation distribution, will allow us to compare behavior at various points in the crystal. It is given simply by:
\begin{equation}f_{\boldsymbol{r}_{0}}^{(L)}(\delta) = g_{\boldsymbol{r}_{0}}^{(L)}\left( {\overline{\varphi}}_{\boldsymbol{r}_{0}}^{(L)} + \delta \right).\end{equation}
While the moments of the orientation distribution \(g_{\boldsymbol{r}_{0}}^{(L)}\) would be given in global coordinates \(\hat{\boldsymbol{b}},\hat{\boldsymbol{a}}\), the moments of the fluctuation distribution \(f\) are given in terms of the local coordinates \({\overline{\boldsymbol{l}}}_{\boldsymbol{r}_{0}},{\overline{\boldsymbol{p}}}_{\boldsymbol{r}_{0}}.\)

Because this now describes deviations from the local average orientation, it is a suitable object to compare across many points in the crystal. For this reason, let us define the global fluctuation distribution for local averages of length \(L\), with points weighted by density:
\begin{equation}
f^{(L)}(\delta) \coloneqq \frac{1}{\left| \mathcal{L} \right|}\int_{\mathcal{M}}^{}{f_{\boldsymbol{r}_{0}}^{(L)}(\delta)\rho\left( \boldsymbol{r} \right)d^{3}\boldsymbol{r}} \label{eq:6globFDFdef}
\end{equation}
where \(\left| \mathcal{L} \right|\) is the total dislocation line length in the crystal, which is also the integral norm of the density. By examining this global fluctuation distribution, we have removed all dependencies save the coarse-graining length. We can thus examine the behavior of this distribution at various coarse-graining lengths and comment on the various approximations which might be appropriate at various scales. As a result, the behavior of this global distribution can be used to investigate regimes of appropriate kinematic theories of dislocation motion.

\subsection{The role of local line orientation distribution in dislocation transport}
The line orientation distribution we have just defined has, in general, its own dynamics. Dislocation transport theories such as vector density dislocation dynamics and multipole dislocation dynamics are reduced forms of these more general dynamics, and so our considerations should begin with the evolution of the dislocation distribution over both space and orientation.

Let us begin with the transport of discrete dislocations. At this fundamental level, the transport equation follows mathematically from the evolution of a system of closed lines under a prescribed velocity field\footnote{This represents the Lie derivative of the dislocation density (a two-form) under the action of the velocity vector field. The full Lie derivative also involves \(\boldsymbol{v\ }\nabla\cdot\boldsymbol{\rho}\), but this is omitted as the divergence term is null for closed curves. For a consideration of open curves, which become relevant in the consideration of dislocation networks, cf.~\cite{Starkey2022}.}:
\begin{align}
    \frac{\partial}{\partial t}\boldsymbol{\rho}&\coloneqq\nabla\times\left( \boldsymbol{v}\times\boldsymbol{\rho} \right),\\
    \boldsymbol{\rho}(\boldsymbol{r}) &=\rho\left( \boldsymbol{r} \right)\boldsymbol{l}\left( \boldsymbol{r} \right).
\end{align}
Note that in its microscopic form, the orientation field is single-valued: every location has only one orientation at this point. Upon averaging, we obtain:
\begin{equation}
\frac{\partial}{\partial t}\left\langle \boldsymbol{\rho} \right\rangle = \nabla \times \left\langle \boldsymbol{v} \times \boldsymbol{\rho} \right\rangle.\label{eq:7AvDynamics}
\end{equation}
We cannot express the average slip rate \(\left\langle \boldsymbol{v} \times \boldsymbol{\rho} \right\rangle\) in terms of the average velocity and density vectors unless all dislocation lines crossing the volume  $\Omega_{\boldsymbol{r}}^{(L)}$ have the same orientation ~\cite{Hochrainer2007,Hochrainer2015}. This generally is not the case if single-slip statistical storage occurs. In such a case, it is obvious that the average transport equation is not `kinematically closed'; that is, the transport of the density vector cannot be expressed in terms of the density vector and a velocity field. As a result, we must seek a more general transport equation to average.

The transport of dislocations in space is elegantly extended by a canonical lifting operation, the development of which we owe to Hochrainer~\cite{Hochrainer2007}. By means of the canonical association between an angular coordinate and a vector in space, we can treat a density distribution not only in space alone, but the direct product of space with an orientation space (e.g., the unit circle $S^{1}$). In fact, this is exactly what we have already defined in the local orientation distribution function {[}Eq. (2){]}. This `higher-order density' is given by:
\begin{equation}\rho_{\text{HO}}\left( \boldsymbol{r},\varphi \right) \coloneqq \rho\left( \boldsymbol{r} \right)g^{(L)}_{\boldsymbol{r}}(\varphi).\label{eq:26_HOdef}\end{equation}
The higher-order line direction now becomes a four-dimensional vector,
linking the orientation coordinate to its canonical direction in space
as well as to the curvature of the lines \(\kappa\). Representing by
\((\boldsymbol{v},a)\ \)a four-vector with spatial part \(\boldsymbol{v}\) and
angular part \(a\), the higher-order line direction becomes:
\begin{equation}
\boldsymbol{L}\left( \boldsymbol{r},\varphi \right) \coloneqq \Big( \boldsymbol{l}(\varphi),\ \kappa\left( \boldsymbol{r},\varphi \right) \Big)
\end{equation}
The higher-order density vector, now including a curvature density $q$, is then given by
\begin{align}
    \boldsymbol{\rho}_\text{HO}\left( \boldsymbol{r},\varphi \right)&=\Big( \rho_\text{HO}\left( \boldsymbol{r},\varphi \right)\boldsymbol{l}(\varphi),q\left( \boldsymbol{r},\varphi \right) \Big)\\
    q\left( \boldsymbol{r},\varphi \right) &\coloneqq \rho_\text{HO}\left( \boldsymbol{r},\varphi \right)\kappa(\boldsymbol{r},\varphi)
\end{align}

The same lifting operation that links the orientation space with the
curvature also specifies the angular portion of the velocity field (which can be thought of as rotating the lines) as
\begin{align}
    \vartheta\left( \boldsymbol{r},\varphi \right) &= \left( \boldsymbol{L}\cdot\nabla^\text{HO} \right)v\left( \boldsymbol{r},\varphi \right)\\
    &= (\boldsymbol{l}(\varphi)\cdot\nabla)v\left( \boldsymbol{r},\varphi \right)+\kappa\left( \boldsymbol{r},\varphi \right)\partial_{\varphi}v\left( \boldsymbol{r},\varphi \right)
\end{align}
where \(\nabla^\text{HO} = \left( \nabla,\partial_{\varphi} \right)\). The
resulting total velocity vector is given by:
\begin{equation}\boldsymbol{V}\left( \boldsymbol{r},\varphi \right) \coloneqq \Big( v\left( \boldsymbol{r},\varphi \right)\boldsymbol{p}(\varphi),\ \vartheta\left( \boldsymbol{r},\varphi \right) \Big).\end{equation}
The higher-order transport equation then follows again mathematically
from the transport of closed curves, yielding a similar form to the
original transport equation:\footnote{The cross product and curl here technically denote, in the language of differential geometry, an exterior product and an exterior derivative followed by a dual operation.}
\begin{equation}
\frac{\partial}{\partial t}\boldsymbol{\rho}^\text{HO}=\nabla^\text{HO}\times\left( \boldsymbol{V} \times\boldsymbol{\rho}^\text{HO} \right).
\end{equation}
While in this form we can see the analogy to the original transport
equation, it is often more instructive in the form of coupled equations
for \(\rho\) and \(q\):
\begin{subequations}\label{eq:8hdcddev}
    \begin{align}
    \partial_{t}\rho &= \left( \boldsymbol{p} \cdot\nabla \right)(\rho v) + \partial_{\varphi}(\rho\vartheta) - qv\\
    \partial_{t}q &= \left( \boldsymbol{p }\cdot\nabla \right)(qv) - \left( \boldsymbol{l }\cdot\nabla \right)(\rho\vartheta)
\end{align}
\end{subequations}

In the original transport equation [Eq. \ref{eq:7AvDynamics}] the requirement for kinematic closure was that all dislocation lines crossing the volume share the same direction: a stringent requirement indeed. The kinematic linking of curvature and the orientation space allows us to kick this requirement further down the line. The requirement for kinematic closure in the higher-order transport equation is simply that dislocations at a point in the higher-order space (i.e. dislocations at the same spatial location and sharing an orientation) must all have identical curvatures. This requirement is generally taken to be more realistic in dislocation systems treated significantly above the discrete length scale.

We see that the higher-order transport equation, which must be considered in systems where the orientation distribution is non-trivial, is intimately associated with the orientation distribution function \(g^{(L)}_{\boldsymbol{r}}(\varphi)\) {[}recall Eq. (\ref{eq:26_HOdef}){]}. Both spatial and directional partial derivatives in Eqs. (\ref{eq:8hdcddev}) will contain terms which reflect the spatial variation of the local orientation distribution in space as well as the shape of the distribution in orientation space.

This theory is not without its drawbacks, however. The price paid for this kinematic consistency is having to consider the full orientation space at every point in the crystal. Because of this, numerical solution of this transport is a complicated problem in its own right~\cite{Sandfeld2010}. The way around this problem is a reduced representation of the local orientation distribution. This reduced representation is found by considering the evolution of various moments of the local orientation distribution called alignment tensors. This representation will provide the context for approximations which bridge the discrete and higher-order theories.

\section{Reduced descriptions of local dislocation
collections\label{sec:3_reduced_descriptions}}

We saw briefly that only the geometrically necessary dislocation content and the average line orientation give rise to long-range mechanical fields in the crystal. It is tempting, then, to consider only the dynamics of the vector density. However, we shall now see that these dynamics are affected by higher moments of the orientation distribution; closure principles for these moments will yield a range of possible approximations between which the global fluctuation distribution can then adjudicate.

The reduced representation can be summarized as follows (cf. \cite{Hochrainer2015,Monavari2016}). Consider the following averages of the density \(\rho\left( \boldsymbol{r},\varphi \right)\)(and similarly, the
curvature density \(q\left( \boldsymbol{r},\varphi \right)\)) over the
angular space, let us call these the tensor moments of the higher-order density field:
\begin{subequations}
\begin{align}
    \rho^{(0)}\left( \boldsymbol{r} \right)&\coloneqq\int_{S^{1}}^{}{\rho\left( \boldsymbol{r},\varphi \right)d\varphi},\\
    &= \rho\left( \boldsymbol{r} \right)\int_{S^{1}}^{}{g_{\boldsymbol{r}}^{(L)}(\varphi)d\varphi}\\
    &= \rho\left( \boldsymbol{r} \right)\\
    \boldsymbol{\rho}^{(1)}\left( \boldsymbol{r} \right) &\coloneqq \rho\left( \boldsymbol{r} \right)\int_{S^{1}}^{}{g_{\boldsymbol{r}}^{(L)}(\varphi)\boldsymbol{l}(\varphi)d\varphi}\\
    &= \rho\left( \boldsymbol{r} \right)\beta_{1}\left( \boldsymbol{r} \right){\overline{\boldsymbol{l}}}_{\boldsymbol{r}}^{\left(L \right)}\label{eq:1circavAltens}\\
    \boldsymbol{\rho}^{\left( \boldsymbol{2} \right)}\left( \boldsymbol{r} \right)&\coloneqq\rho\left( \boldsymbol{r} \right)\int_{S^{1}}^{}{g_{\boldsymbol{r}}^{(L)}(\varphi)\boldsymbol{l}^{\otimes 2}(\varphi)d\varphi}\\
    &\qquad\vdots\nonumber\\
    \boldsymbol{\rho}^{(n)}\left( \boldsymbol{r} \right)&\coloneqq\rho\left( \boldsymbol{r} \right)\int_{S^{1}}^{}{g_{\boldsymbol{r}}^{(L)}(\varphi)\boldsymbol{l}^{\otimes n}(\varphi)d\varphi}
\end{align}
\end{subequations}
where the tensor power is defined simply by repeated application of the
tensor power. For a vector \(\boldsymbol{u}\) we can define it recursively
\(\boldsymbol{u}^{\otimes n}=\boldsymbol{u}^{\otimes n - 1}\boldsymbol{\otimes u}\). Removing the zeroth order component we arrive at simply an alternative formulation of the orientation distribution function in terms of these tensor moments; because these tensor moments capture various degrees of alignedness of the collection these are termed the alignment tensors $\mathbf{T}^{(n)}(\boldsymbol{r})$, defined as 
\begin{equation}
\boldsymbol{\rho}^{(n)}\left( \boldsymbol{r} \right)=\rho\left( \boldsymbol{r} \right)\mathbf{T}^{(n)}\left( \boldsymbol{r} \right),
\end{equation}
where
\begin{equation}
\mathbf{T}^{(n)}\left( \boldsymbol{r} \right)\coloneqq\left\langle \boldsymbol{l}^{\otimes n}(\varphi) \right\rangle_{\boldsymbol{r}}^{(L)}\label{eq:10alignDef}
\end{equation}
Notice that we have already considered the `first circular average' in Eq. (\ref{eq:4avorient}) and the first alignment tensor is related to it rather simply in Eq. (\ref{eq:1circavAltens}). These alignment tensors are related to the higher-order moments of the orientation distribution.  

The reduced dynamics follows primarily from the linearity of the integral, but also due to the following relations
between derivatives of the unit tangent:
\begin{subequations}
\begin{align}
    \partial_{\varphi}\ \boldsymbol{l}(\varphi) &= \boldsymbol{p}(\varphi),\\
    \partial_{\varphi}\ \boldsymbol{p}(\varphi) &= - \boldsymbol{l}(\varphi).
\end{align}
\end{subequations}
Using these properties, the high-dimensional evolution equations \cref{eq:8hdcddev} can be used to derive
evolution equations for the alignment tensors, given below:
\begin{subequations}\label{eq:11_reducedEvolution}
\begin{align}
    \partial_{t}\rho^{(0)} =& \nabla \cdot \left( v\boldsymbol{\ }\hat{\boldsymbol{n}}\times\boldsymbol{\rho}^{\left(1 \right)} \right)+vq^{(0)}\\
    \partial_{t}\boldsymbol{\rho}^{\left( \boldsymbol{n} \right)}=&\text{sym}\Bigg\lbrack \nabla\times\left( v\ \hat{\boldsymbol{n}}\otimes \boldsymbol{\rho}^{\left( n - 1 \right)} \right)\nonumber\\
    &+(n - 1)v\mathbf{Q}^{(n)} \nonumber\\
    &- (n - 1)\left( \hat{\boldsymbol{n}}\times\boldsymbol{\rho}^{(n + 1)} \right) \cdot \nabla v \Bigg\rbrack\\
    \partial_{t}q^{(0)} =& \nabla \cdot \left( v\mathbf{Q}^{(1)} - \boldsymbol{\rho}^{(2)}\cdot\nabla v \right).
\end{align}
\end{subequations}
where $\text{sym}[\mathbf{A}]$ represents symmetrization over all tensor indices and \(\mathbf{Q}^{(n)}\) are the auxiliary curvature tensors of
the form:
\begin{equation}
\mathbf{Q}^{(n)}=\int_{}^{}{q\left( \boldsymbol{r},\varphi \right)\ \boldsymbol{p}^{\otimes 2}(\varphi)\otimes \boldsymbol{l}^{\otimes n - 2}(\varphi)}\boldsymbol{\ }d\varphi.
\end{equation}
For \(n \leq 2\) the power of \(\boldsymbol{p}\) is equal to \(n\). Notice
that each of the evolution equations in Eqs. (\ref{eq:11_reducedEvolution}) is coupled to the next equation in the hierarchy:
\(\partial_{t}\boldsymbol{\rho}^{(n)}\) involves a \(\boldsymbol{\rho}^{(n + 1)}\) term. Unless we
close the hierarchy at a low order, we have gained nothing: we have just
changed the manner in which the full orientation space is represented.

In order to close the hierarchy at order \(m\), it is necessary to
determine some closure function $h$ by which:
\begin{equation}
\mathbf{T}^{(m + 1)}(\boldsymbol{r})=h\left( \mathbf{T}^{(1)},\ldots,\mathbf{T}^{(m)} \right)\label{eq:closure}
\end{equation}
Notice that this closure relation must hold at all points
\(\boldsymbol{r}\). As a result, we expect the appropriateness of a
closure relation to depend only on the coarse-graining length \((L)\)
considered. Before proceeding, let us consider two differing (but equivalent) representations of the alignment tensors \cref{eq:10alignDef} in terms of the symmetric and antisymmetric moment fields:
\begin{subequations}
\begin{align}
M_{n}^{(L)}\left( \boldsymbol{r} \right)&\coloneqq\left\langle \cos^{n}\delta \right\rangle_{\boldsymbol{r}}^{(L)}\\
    {\widetilde{M}}_{n}^{(L)}\left( \boldsymbol{r} \right)&\coloneqq\begin{dcases}
\left\langle \sin^{n}\delta \right\rangle_{\boldsymbol{r}}^{(L)} & n\text{ odd} \\
\left\langle \cos\delta\sin^{n - 1}\delta \right\rangle_{\boldsymbol{r}}^{(L)} & n\text{ even}
\end{dcases}
\end{align}
\end{subequations}
or the characteristic sequence of the fluctuation distribution
\begin{equation}
\beta_{n}^{(L)}\left( \boldsymbol{r} \right) \coloneqq \left\langle z^{n}(\delta) \right\rangle_{\boldsymbol{r}}^{(L)}.
\end{equation}
For the remainder of the present work, we assume that the fluctuation distribution is symmetric, resulting in the characteristic sequence being strictly real (and thus symmetric upon reversal of the sign of the index) and that all antisymmetric moments are zero:
\begin{subequations}
\begin{align}
    \beta_{- |k|} &= \beta_{|k|},\\
    {\widetilde{M}}_{k} &= 0.
\end{align}
\end{subequations}Lastly, we state the form of the second and third order alignment
tensors (cf. Appendix A):
\begin{subequations}
\begin{align}
    \left\langle \boldsymbol{l}^{\otimes 2}(\delta) \right\rangle &= \frac{1}{2}\left\lbrack \left( 1 + \beta_{2} \right)\overline{\boldsymbol{ll}}+\left( 1 - \beta_{2} \right)\overline{\boldsymbol{pp}} \right\rbrack\\
    \left\langle \boldsymbol{l}^{\otimes 3}(\delta) \right\rangle &= \frac{1}{4}\left\lbrack \left( 3\beta_{1} + \beta_{3} \right)\overline{\boldsymbol{lll}}+\left( 3\beta_{1} - \beta_{3} \right)\overline{\boldsymbol{lpp}} \right\rbrack,
\end{align}\label{eq:alignmentTensorsChar}\end{subequations}
where $\overline{\boldsymbol{lpp}}= \text{sym}[\overline{\boldsymbol{l}}\otimes \overline{\boldsymbol{p}}^{\otimes 2}]$. In general, the closure problem which we will now consider can equivalently be phrased in the alignment tensor hierarchy [Eq. (\ref{eq:closure})] or in the $\beta_k$ hierarchy or the $M_k$. With this in mind, let us consider two candidate closure relations: line bundle closure---a contribution of the present work---and the maximum entropy closure, which is the current method in use \cite{Monavari2016}.

\subsection{The line bundle expansion of the alignment tensor
hierarchy}

The line bundle expansion which we will present in this section is
motivated by the following observation. At very small length scales, the dislocations in a collection near a point are all parallel.\footnote{In the extreme case of small length $L_0$, only a single dislocation can be present at the point.} The result is
that the fluctuation distribution is a Dirac mass\footnote{We apologize
  that the notation \(\delta\) is overloaded. To distinguish the Dirac
  delta from the fluctuation coordinate, we denote the former by
  \(\hat{\delta}\).} and the following relations hold for the
characteristic sequence and alignment tensors:
\begin{subequations}
\begin{align}
    f_{\boldsymbol{r}}^{\left( L_{0} \right)}(\delta) &= \hat{\delta}(\delta)\\
    \beta_{k}^{\left( L_{0} \right)}\left( \boldsymbol{r} \right) &= 1\qquad \forall k \in \mathbb{Z}\\
    \left\langle \boldsymbol{l}^{\otimes n}(\delta) \right\rangle_{\boldsymbol{r}}^{\left( L_{0} \right)}&=\overline{\boldsymbol{l}}_{\boldsymbol{r}}^{ \ \otimes n}\label{eq:12_dirac_alignment}
\end{align}
\end{subequations}
Well, what relations between alignment tensors would obtain if we go to slightly longer length scales? Here we might be including one or more bowed dislocations which form a `smooth line bundle' \cite{Sedlacek2010,Lin2021}. In this regime, the polarization would differ from unity, but only slightly: 
\begin{equation}
\beta_{1}^{(L)}\left( \boldsymbol{r} \right) = 1 - \epsilon\left( \boldsymbol{r} \right),\qquad 0 < \epsilon \ll 1\end{equation}
and the angular distribution would be sharply focused around zero fluctuation. Inspired by the discrete relation for the alignment tensor as simply the tensor power of the average direction \cref{eq:12_dirac_alignment}, we propose a more modest goal, that we might factor out of the $n$-th alignment tensor a single factor of \(\overline{\boldsymbol{l}}\):

\begin{equation}\left\langle \boldsymbol{l}^{\otimes n}(\delta) \right\rangle_{\boldsymbol{r}}^{(L)} \approx \text{sym}\left\lbrack {\overline{\boldsymbol{l}}}_{r} \otimes \left\langle \boldsymbol{l}^{\otimes n}(\delta) \right\rangle_{\boldsymbol{r}}^{(L)} \right\rbrack.\ \ \end{equation}
In order to perform this factoring, what closure relation would need to obtain in the characteristic sequence? For how low of polarization would this relation hold? The true closure would take the form of a ladder operator for the characteristic sequence. Without loss of generality, we may define a ladder operator \(\eta_{k}^{\pm}\) as the ratio between neighboring entries in the characteristic sequence: 
\begin{equation}
\eta_{k}^{\pm}\beta_{k} = \beta_{k \pm 1}.
\end{equation}
The approximation we make is to consider the fluctuation distribution to be `Cauchy-like', having the ladder operator:
\begin{subequations}
\begin{align}
    \eta^+_k &\approx \begin{dcases}
        \beta_1 &k\geq 0\\
        \beta_1^{-1} &k <0 \\ 
    \end{dcases}\\
    \eta^-_k &\approx \begin{dcases}
        \beta_1^{-1} &k> 0\\
        \beta_1  &k \leq0 \\ 
    \end{dcases}
\end{align}
\label{eq:13_ladder}
\end{subequations}
Such a distribution we term `Cauchy-like' because this relation would
hold exactly for a (wrapped) Cauchy distribution \cite{mardiaStatisticsDirectionalData1975}, which has a geometric
characteristic sequence:
\begin{equation}
\beta_{k}^{(\text{Cauchy})} = \beta_{1}^{|k|}.
\end{equation}

The definition of a closure approximation in terms of a ladder operator keeps us from over committing to the Cauchy shape; the closure at \(n\)-th order simply requires:
\begin{equation}
\eta_{n}^{+}\beta_{n} = \beta_{n + 1}.
\end{equation}
What are the effects of this approximation to the characteristic
sequence? Closing the sequence in this manner results in the following
expansion of the \(n\)th order alignment tensor:

\begin{equation}\left\langle \boldsymbol{l}^{\otimes n}(\delta) \right\rangle_{\boldsymbol{r}}^{(L)} \approx sym\left\lbrack {\overline{\boldsymbol{l}}}_{r} \otimes \left\langle \boldsymbol{l}^{\otimes n}(\delta) \right\rangle_{\boldsymbol{r}}^{(L)} \right\rbrack + O(\epsilon).\end{equation}

For a detailed derivation of these relations by means of this recursion
relation for the Fourier components of the tensor power sequence shown
in Eq. (11), we refer the reader to Appendix A.

In order to test such an assumption, we may evaluate the accuracy of the
following closure approximations at second and third order,
respectively:
\begin{align}
\beta_{2} &\approx \beta_{1}^{2}\label{eq:LB1}\\
\beta_{3} &\approx \beta_1\beta_2\label{eq:LB2}
\end{align}
which correspond to the following expressions for the alignment tensors [cf. Eq. (\ref{eq:alignmentTensorsChar})]:
\begin{subequations}
\begin{align}
    \left\langle \boldsymbol{l}^{\otimes 2}(\delta) \right\rangle_{\boldsymbol{r}}^{(L)}&\approx\frac{1}{2}\left\lbrack \left( 1 + \beta_{1}^{2} \right)\overline{\boldsymbol{ll}} + \left( 1 - \beta_{1}^{2} \right)\overline{\boldsymbol{pp}} \right\rbrack,\\
    \left\langle \boldsymbol{l}^{\otimes 3}(\delta) \right\rangle_{\boldsymbol{r}}^{(L)}&\approx\frac{\beta_{1}}{4}\left\lbrack \left( 3 + \beta_{2} \right)\overline{\boldsymbol{ll}} + \left( 2\beta_{1}^{- 2} + 1 - \beta_{2} \right)\overline{\boldsymbol{pp}} \right\rbrack.
\end{align}
\end{subequations}

\subsection{Maximum entropy closure of the alignment tensor
hierarchy}
For comparison, we will also present here the closure approximations for the maximum entropy approach due to Monavari et al. \cite{Monavari2016}. The maximum entropy approach derives from finding the extremum of the entropy functional
\begin{equation}S\lbrack f\rbrack \coloneqq \int_{S^{1}}^{}{f(\delta)\ln{f(\delta)}d\delta}\end{equation}
subject to the constraint of the first \(n\) moments of the
distribution:
\begin{equation}
\int_{S^{1}}^{}{f(\delta)\cos^{k}\delta} = M_{k},\qquad k = 1,\ldots n.
\end{equation}
This common approach formalizes a concept of the `maximally ignorant' distribution which could agree with the first \emph{n} known moments. The result is that the fluctuation distribution can be given in terms of Lagrange multipliers $\lambda_{1}, \dots, \lambda_n$, which in turn are functions of the known moments:
\begin{equation}
f(\delta) \approx \frac{1}{Z\left( M_{1},\ldots,M_{n} \right)}\exp{\left\lbrack {\sum_{k = 1}^{n}{\lambda_{k}\left( M_{1},\ldots M_{n} \right)}\cos^{k}}\delta \right\rbrack.}
\end{equation}
$Z$ is the partition function, serving to normalize the distribution. From this approximate distribution, the higher order moments can be evaluated. The Lagrange parameters tend to be non-linear functions of the constrained moments. The resulting closure relations for the higher order moment functions are closely approximated in the relevant regions of the space of lower order moments by the following polynomials. For closure at second order, the second moment is given approximately by \cite{Monavari2016}:
\begin{equation}
M_{2}^{(L)}(\boldsymbol{r)} = \frac{1}{4}\left( 2 + M_{1}^{2} + \beta_{1}^{6} \right).
\end{equation}
For closure at third order, the third moment is given by:
\begin{equation}M_{3}^{(L)}(\boldsymbol{r)} = M_{1}\sqrt{M_{2}}.\end{equation}
By utilizing the power reduction formulae describing the equivalence between the characteristic sequence and moments, these reduce, respectively, to the following closure approximations for the characteristic sequence:
\begin{align}
\beta_{2}^{(L)}\left( \boldsymbol{r} \right) &\approx \frac{1}{2}\left( \beta_1^2 + \beta_{1}^{6} \right),   \label{eq:ME1} \\
\beta_{3}^{(L)}\left( \boldsymbol{r} \right) &\approx 2\sqrt{2}\beta_{1}\left( \sqrt{1 + \beta_{2}} - 3 \right).\label{eq:ME2}
\end{align}
Compared to the line bundle closure, the maximum entropy closure more aggressively flattens the orientation distribution, as is seen from the $\beta_1^6$ term in the first order closure.

\subsection{A note on the closure of local and global orientation
distributions}

We now have two sets of closure relations, the line bundle relations [\cref{eq:LB1,eq:LB2}] and the maximum entropy relations [\cref{eq:ME1,eq:ME2}]. There is one remaining issue which must be addressed before we can evaluate the closure relations from discrete dislocation data. This issue is that the closure relations are all relations between local moments of the fluctuation distribution: the characteristic sequence in general may vary from point to point within the crystal. As we will now show, the requirement for the closure relations to obtain with respect to the global distribution [\cref{eq:6globFDFdef}] is not that the fluctuations are identically distributed, but only that they are independently distributed. Thus, while the local characteristic sequence may differ from the global characteristic sequence, the closure relations which obtain locally also obtain globally.

It is clear that the global average of a local angular average is equal to the equivalent angular average with respect to the global fluctuation distribution, regardlesss of statistical independence. Letting the following bracket notation denote the spatial average of some quantity $C\left( \boldsymbol{r}\right)$ over the domain $\mathcal{M}$:
\begin{equation}
    \left\lbrack C\left( \boldsymbol{r} \right) \right\rbrack_{\mathcal{M}}^{(L)} \coloneqq \frac{1}{\left| \mathcal{L} \right|}\int_{\mathcal{M}}^{}\rho^{(L)}\left( \boldsymbol{r} \right)C\left( \boldsymbol{r} \right)d^{3}\boldsymbol{r},
\end{equation}
the statement above follows simply by interchange of integrals:
\begin{subequations}
\begin{align}
    \left\langle A(\delta) \right\rangle^{(L)} &= \int_{S^{1}}^{}{d\delta\ A(\delta)}f^{(L)}(\delta)\\
    &= \int_{S^{1}}^{}{d\delta\ A(\delta)}\left\lbrack f_{\boldsymbol{r}}^{(L)}(\delta) \right\rbrack_{\mathcal{M}}^{(L)}\\
    &= \left\lbrack \int_{S^{1}}^{}{d\delta\ A(\delta)f_{\boldsymbol{r}}^{(L)}(\delta)} \right\rbrack_{\mathcal{M}}^{(L)}\\
    &= \left\lbrack \left\langle A(\delta) \right\rangle_{\boldsymbol{r}}^{(L)} \right\rbrack_{\mathcal{M}}^{(L)}.
\end{align}\label{eq:interchangeofav}
\end{subequations}

The above simple case would extend to any linear combination of angular averages as well. However, all of the closure relations involve non-linear functions of angular averages [Eqs. (\ref{eq:LB1}-\ref{eq:LB2}, \ref{eq:ME1}-\ref{eq:ME2})]. That these relations would hold in the case of the global fluctuation distribution is not immediately clear. In order to show that the non-linear functions of angular averages (e.g. \(\beta_{1}^{2}\), \(\sqrt{1 + \beta_{2}}\), \(\beta_{1}\beta_{2}\)) can be evaluated from the global distribution, we simply wish to show that:
\begin{equation}
\left\lbrack \left\langle A(\delta) \right\rangle_{\boldsymbol{r}}^{(L)}\left\langle B(\delta) \right\rangle_{\boldsymbol{r}}^{(L)} \right\rbrack_{\mathcal{M}}^{(L)} = \left\langle A(\delta) \right\rangle^{(L)}\left\langle B(\delta) \right\rangle^{(L)}. \label{eq:18_loctoglobclos}
\end{equation}

By introducing a second volume average over \(\boldsymbol{r}'\), we may rewrite this equation in such a way that the issue is clearer:
\begin{widetext}
\begin{equation}\left\lbrack \left\lbrack \left\langle A(\delta) \right\rangle_{\boldsymbol{r}}^{(L)}\left\langle B(\delta) \right\rangle_{\boldsymbol{r}^{\boldsymbol{'}}}^{(L)}\hat{\delta}\left( \boldsymbol{r -}\boldsymbol{r}^{\boldsymbol{'}} \right) \right\rbrack_{\mathcal{M}}^{(L)} \right\rbrack_{\mathcal{M}^{'}}^{(L)} = \left\lbrack \left\lbrack \left\langle A(\delta) \right\rangle_{\boldsymbol{r}}^{(L)}\left\langle B(\delta) \right\rangle_{\boldsymbol{r}^{\boldsymbol{'}}}^{(L)} \right\rbrack_{\mathcal{M}}^{(L)} \right\rbrack_{\mathcal{M}^{'}}^{(L)}
\end{equation}
\end{widetext}
The requirement for this equality to hold is that spatial measurements
of the fluctuation distribution must be independent. We assume the
fluctuation distributions are statistically independent,\footnote{An
  assumption supported by the data as calculated in the following
  section.} and thus the second integration over \(\boldsymbol{r}'\) is
unaffected by the fact that it must consider the same field point as the
first integration over \(\boldsymbol{r}\).

Because the products of local angular averages can be represented as
products of the equivalent global average [\cref{eq:18_loctoglobclos}], and assuming that the closure relation $h$ has a Taylor expansion, we see that:
\begin{align}\bigg\lbrack h\Big( \left\langle A(\delta) \right\rangle_{\boldsymbol{r}}^{(L)},&\left\langle B(\delta) \right\rangle_{\boldsymbol{r}}^{(L)},\ldots \Big) \bigg\rbrack_{\mathcal{M}}^{(L)} = \nonumber\\
 & h\left( \left\langle A(\delta) \right\rangle^{(L)},\left\langle B(\delta) \right\rangle^{(L)},\ldots \right)
\end{align}

Thus, it suffices for our evaluation of the closure relations [Eqs. (\ref{eq:LB1}-\ref{eq:LB2}, \ref{eq:ME1}-\ref{eq:ME2})] to consider their efficacy on the characteristic sequence of the global distribution [\cref{eq:6globFDFdef}] across a spectrum of coarse-graining lengths \(L\).

\section{Calculation of global fluctuation
distributions\label{sec:4_calculation}}

The calculation of the fluctuation distributions from discrete dislocation configurations follows closely the analysis of section II. Some of the quantities left open are now given particular instantiation, and some of the continuous quantities must now be discretized. To begin with, the dislocation arrangement \(\mathcal{L}\) is now represented as the disjoint union of \(N_{\ell}\) discrete segments:
\begin{equation}
\mathcal{L}=\bigcup_{i = 0}^{N_{\ell}}\ell_{i}.
\end{equation}
The crystal space, which is a rectangular domain of size
\(\left( D_{1},D_{2},D_{3} \right)\), is then discretized into a
cubic mesh which divides the longest dimension of the crystal
(\(D_{1}\)) into \(N_{V}\) equal parts:
\begin{equation}
\boldsymbol{r}_{ijk}=L\left( i\hat{\boldsymbol{x}}+j\hat{\boldsymbol{y}} + k\hat{\boldsymbol{z}} \right).\qquad (i,j,k) \in \mathbb{N}^{3}
\end{equation}
for $\hat{\boldsymbol{x}}, \hat{\boldsymbol{y}}, \hat{\boldsymbol{z}}$ are the unit directions in space and the coarse-graining length is given as \(L \coloneqq D_{1}/N_{V}\). The
weight function by which the scalar dislocation density field is defined
at each point is chosen to be the cloud-in-cell function \cite{Birdsall1969}:
\begin{equation}
w_{L}\left( \boldsymbol{\Delta}\boldsymbol{r} \right)\coloneqq \begin{dcases}
\frac{1}{L^{3}}\prod_{I = 1}^{3}\left( 1 - \left| \frac{\Delta r_{I}}{L} \right| \right) & \text{if all } \left| \Delta r_{I} \right| \leq L \\
0 & \text{otherwise}.
\end{dcases}.
\end{equation}
The coarse-graining volume \(\Omega_{ijk}^{(L)}\) is thus the cube of
side length \(2L\) centered on the point \(\boldsymbol{r}_{ijk}\).

Focusing our attention on the collection of dislocation segments
surrounding a point
\(\Lambda_{ijk\ }^{(L)}=\mathcal{L}\cap\Omega_{ijk}^{(L)},\) we see
that we can assign to each segment \(\ell\) in this collection a
weight:
\begin{equation}
w_{\ell}^{(L)} = \int_{\ell}^{}{w_{L}\left( \boldsymbol{r -}\boldsymbol{r}_{ijk} \right)dl}.
\end{equation}
The density at this point is then given by the direct sum of these
weights:
\begin{equation}
\rho\left( \boldsymbol{r}_{ijk} \right)=\sum_{ \ell\in\Lambda_{ijk}^{(L)}}^{}w_{\ell}^{(L)}.
\end{equation}
Since each segment has a well-defined tangent direction
\({\hat{\boldsymbol{l}}}_{\ell}\) and thus orientation angle
\(\varphi_{\ell}\), we can
also define the average direction:
\begin{equation}
\beta_{1}^{(L)}\left( \boldsymbol{r}_{ijk} \right)e^{i{\overline{\varphi}}_{ijk}} = \frac{1}{\rho\left( \boldsymbol{r}_{ijk} \right)}\sum_{\ell \in \Lambda_{ijk}^{(L)}}^{}{w_{\ell}^{(L)}e^{i\varphi_{\ell}}}.
\end{equation}
This allows us to define the fluctuation angle
\(\delta_{\ell} = \varphi_{\ell} - {\overline{\varphi}}_{ijk}\)
for each segment.

These discrete definitions now allow us to define the local and global
fluctuation distributions. Obviously, we cannot consider the continuous
probability distributions. Instead, we will consider the probability that
the fluctuation directions fall in a sequence of \(N_{I}\) small
intervals which partition the unit circle:
\begin{subequations}
\begin{align}
    f_{ijk}^{(L)}\left( \delta \in I_{q} \right) &= \frac{1}{\rho\left( \boldsymbol{r}_{ijk} \right)}\sum_{\ell \in\Lambda_{ijk}^{(L)}}^{}{w_{\ell}^{(L)}1_{I_{q}}\left( \delta_{\ell} \right)},\\
     I_{q} &= \frac{2\pi}{N_{I}}\left\lbrack q - \frac{1}{2},q + \frac{1}{2} \right).
\end{align}
\end{subequations}
In the limit where \(N_{I} \rightarrow \infty\), this discretized
distribution converges to the fluctuation distribution function. We will
consider \(N_{I} = 360\), or single degree intervals. Regardless, the
global distribution can likewise be defined:
\begin{equation}
f^{(L)}\left( \delta \in I_{q} \right) = \frac{1}{\left| \mathcal{L} \right|}\sum_{ijk}^{}{\rho\left( \boldsymbol{r}_{ijk} \right)f_{ijk}^{(L)}\left( \delta \in I_{q} \right)}.
\end{equation}
Global averages over the angular fluctuations can be calculated either
by volume average of the local averages:
\begin{equation}\left\langle A(\delta) \right\rangle^{(L)} = \frac{1}{\left| \mathcal{L} \right|}\sum_{ijk}^{}{\sum_{\ell\in\Lambda_{ijk}^{(L)}}^{}{w_{\ell}^{(L)}A\left( \delta_{\ell} \right)}}\end{equation}
or by means of the discrete distribution function:
\begin{equation}\left\langle A(\delta) \right\rangle^{(L)} = \sum_{q = 1}^{2\pi}{f^{(L)}\left( \delta \in I_{q} \right)A\left( \frac{2\pi q}{N_{I}} \right)}.\end{equation}
Although the errors associated with the discretization of the angular
fluctuation space in the latter expression are small, we will still
prefer to utilize the former.

Now, given a discrete dislocation configuration, we can analyze all of
the angular statistics discussed at various coarse-graining lengths. The
particular discrete dislocation configurations used were obtained from a
set of 45 simulations performed in microMegas \cite{Devincre2011} which we have
utilized previously \cite{Anderson2021}. The simulations considered 4.40 x 4.87 x
\SI{5.74}{\micro\metre} domains of copper, seeded with random dipolar loops
resulting in an initial density of \SI{2}{\micro\metre^{-2}}. These domains were
then deformed at a {constant tensile strain rate along the [001] direction} to 0.3\% plastic strain. {Cross-slip was enabled at 300 K.} 
Dislocation configurations were extracted at strain increments of 0.075\%,
resulting in 180 configurations which will be analyzed according to the
framework just described.

{MicroMegas is a lattice-based discrete dislocation dynamics code, and so segments of a given slip system lie only along 8 discrete directions, which would obviously hinder analysis of orientation statistics. Because of this, glissile segment chains are resampled to segment midpoints to more smoothly represent the curved dislocation with straight segments. This also has the effect of allowing a broader range of segment orientations.}\footnote{{The resampled orientations are still limited to relatively low-order rational directions due to the lattice-based nature of the segment positions.}}{ For more information on this resampling, cf. chapter 5 of \cite{andersonStatisticalFoundationsLine2023}}.

\begin{figure}
    \centering
    \subfloat{\includegraphics[height=2.5in]{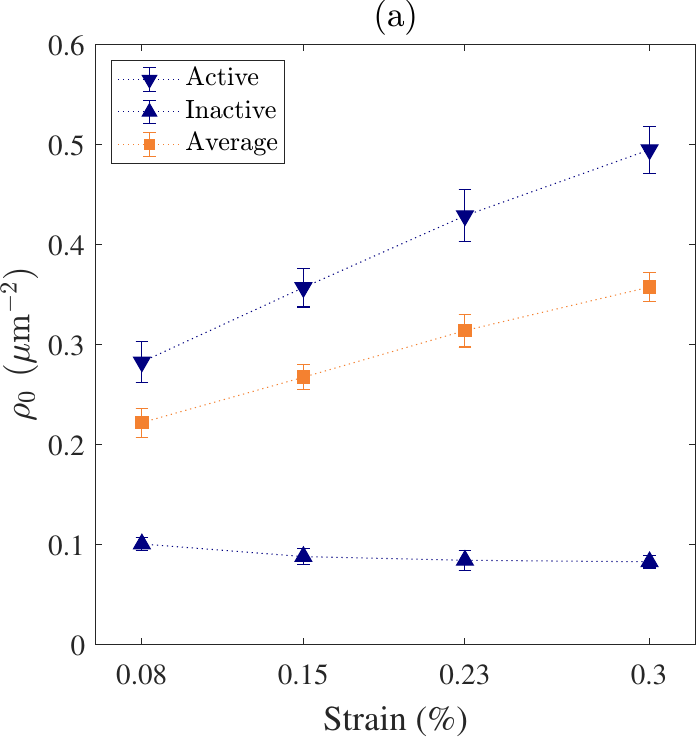}}\\
    \subfloat{\includegraphics[height=2.5in]{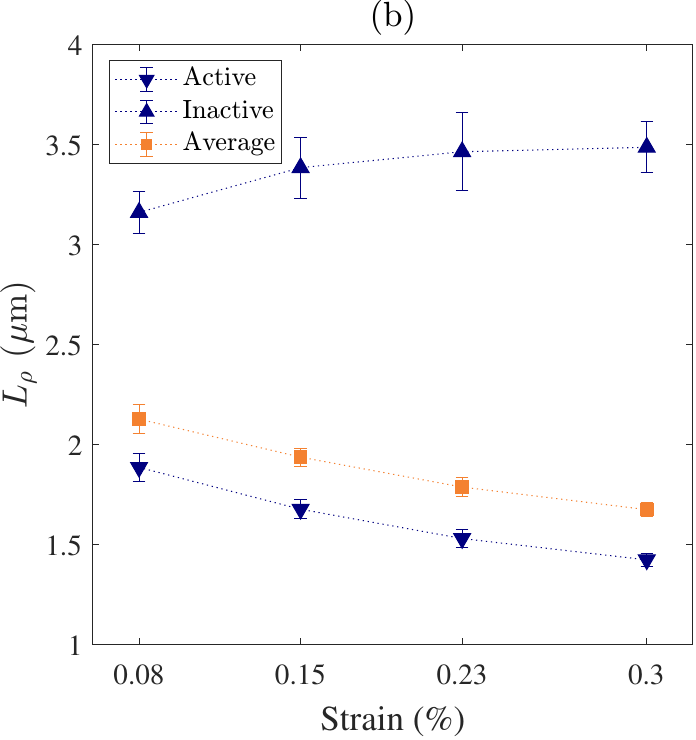}}
    \caption{Typical dislocation densities and mean dislocation spacing from the discrete dislocation configurations. Because of the asymmetry of dislocation multiplication and thus dislocation density on active and inactive slip systems, the calculations of the densities and corresponding average spacings are shown for the cases where these calculations are restricted to active and inactive slip systems as well as the unrestricted average. Error bars represent one standard deviation (over the set of 45 simulations).}
    \label{fig:SimulationStuff}
\end{figure}

The coarse-graining lengths which we will consider range from 71.5 nm to \SI{1.15}{\micro\metre}, corresponding to divisions of the longest edge of the crystal domain into 5 to 80 equal parts. However, we will not analyze these lengths as absolute distances. Rather, we will consider these coarse-graining lengths relative to the average dislocation spacing in the crystal. Because we treat only glissile dislocations and we treat every slip system as having its own fluctuation distribution, the relevant density for determining the average dislocation spacings (\(L_{\rho_{0}} = \left( \rho_{0} \right)^{- 1/2}\)) are:
\begin{equation}\rho_{0} = \frac{1}{12}\sum_{i = 1}^{12}\left| \mathcal{L}_{g}^{\lbrack\alpha\rbrack} \right|\end{equation}
where \(\left| \mathcal{L}_{g}^{\lbrack\alpha\rbrack} \right|\ \)is the total line length of glide dislocations on slip system \(\lbrack\alpha\rbrack\). The resulting dislocation densities and corresponding dislocation spacings are shown in \cref{fig:SimulationStuff}. Because each configuration has a distinct mean dislocation spacing, the absolute coarse-graining lengths will have various values when measured relative to these various dislocation spacings.

With all this in hand, we may now examine the global fluctuation distributions for various coarse-graining lengths. Also, we will compare the average characteristic sequence: 
\begin{equation} \beta_{n}^{(L)} = \left\langle e^{in\delta} \right\rangle^{(L)} \end{equation} 
across these coarse-graining lengths to determine regimes of coarse-graining length where the line bundle or maximum entropy closure relations [Eqs. (\ref{eq:LB1}-\ref{eq:LB2}, \ref{eq:ME1}-\ref{eq:ME2})] are appropriate.

\section{Results \label{sec:5_results}}
\begin{figure}
    \centering
    \subfloat{\includegraphics[height=2.5in]{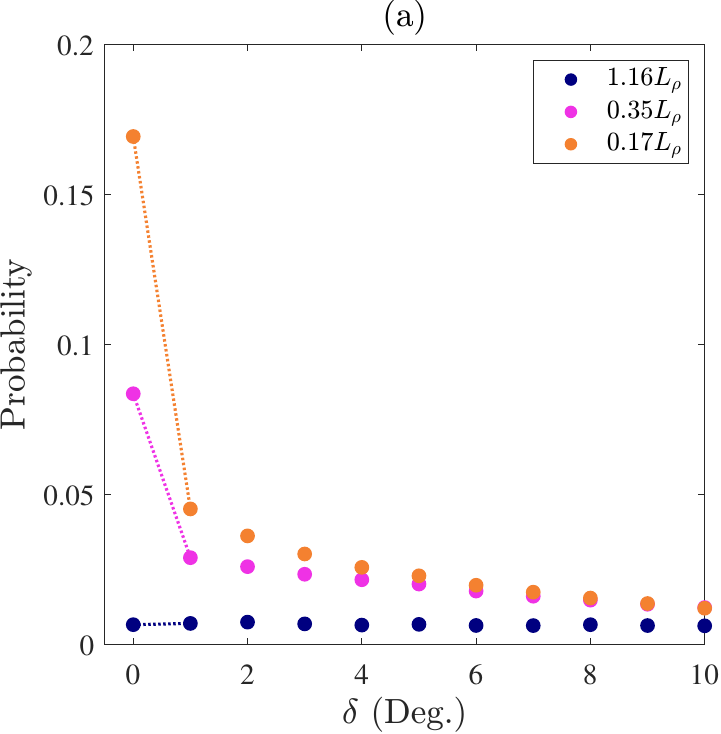}}\\
    \subfloat{\includegraphics[height=2.5in]{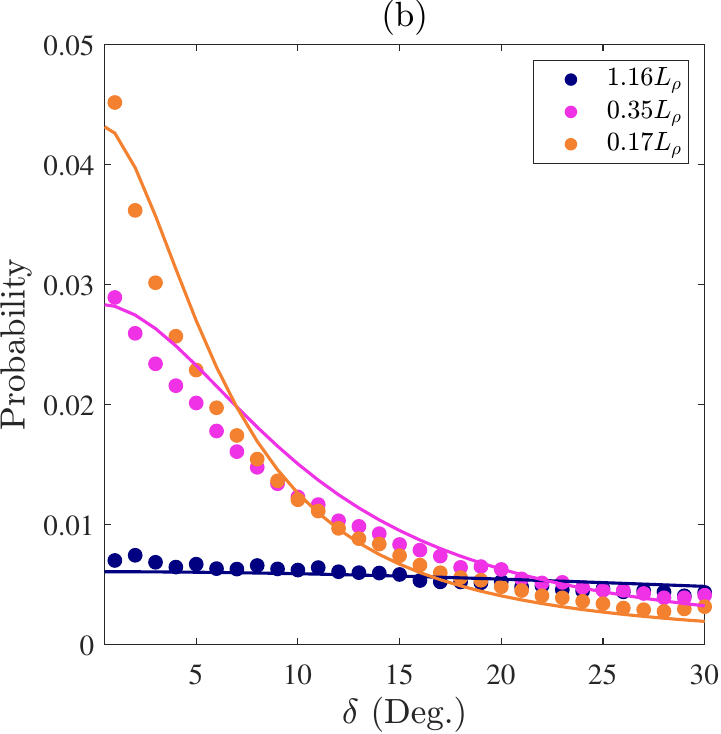}}
    \caption{Typical global orientation fluctuation distributions from the dataset. Both figures show a selection of the orientation fluctuation distribution functions at various coarse-graining lengths. Fig. 2a shows the sharp peak at zero fluctuation angle coming from single-crossing events, and Fig. 2b shows the same functions with the zero fluctuation point shielded from view. Cauchy fits are shown on the latter figure to demonstrate the shape of the distributions at non-zero fluctuation angle.}
    \label{fig:Distributions}
\end{figure}

While it is the global characteristic sequence that is most relevant to the closure relations this work has set out to evaluate, it is nonetheless instructive to see the shape of the global distributions themselves. We would like to remind the reader that these distributions do not hold at every local point in the crystal--the local distributions are poorly behaved due to the discrete nature of the segments--they do describe the average strength of orientation fluctuations in the crystal. Let us turn, then, our attention to these global distributions, shown in \cref{fig:Distributions}. It is noted that there is a sharp peak at zero fluctuation (\cref{fig:Distributions}a), corresponding to the local regions being pierced only by a single dislocation. In these cases, the local fluctuation distribution is a Dirac mass at zero fluctuation. The strength of this peak in the global distribution can be thought of as the density-weighted fraction of such regions in the crystal. The remaining distribution at non-zero fluctuations (\cref{fig:Distributions}b), is closer in shape to a Cauchy distribution than to a von Mises distribution. The general trend---as can be seen from \cref{fig:Distributions}---is that short coarse-graining lengths correspond to narrow distributions about zero fluctuation, and long coarse-graining lengths to broad distributions.

\begin{figure}
    \centering
    \subfloat{\includegraphics[height=2.5in]{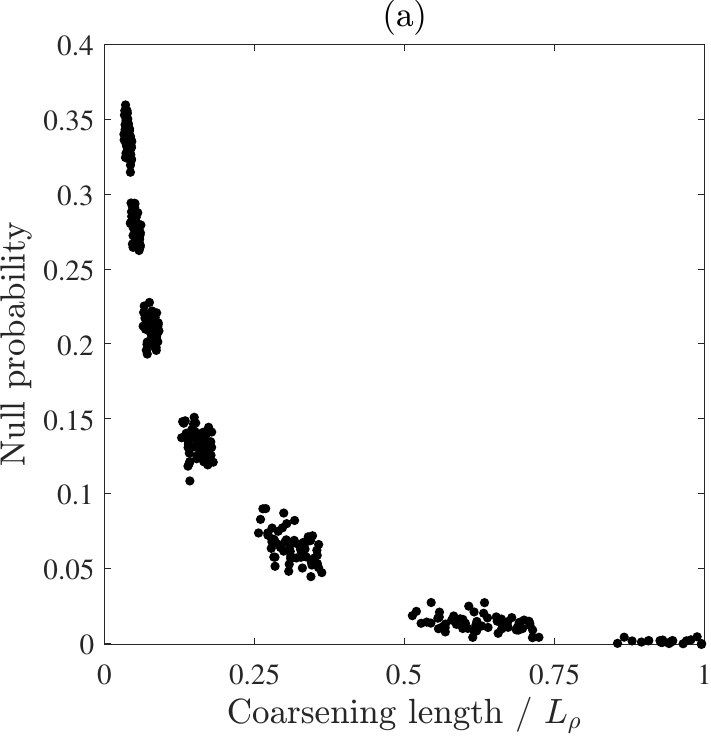}}\\
    \subfloat{\includegraphics[height=2.5in]{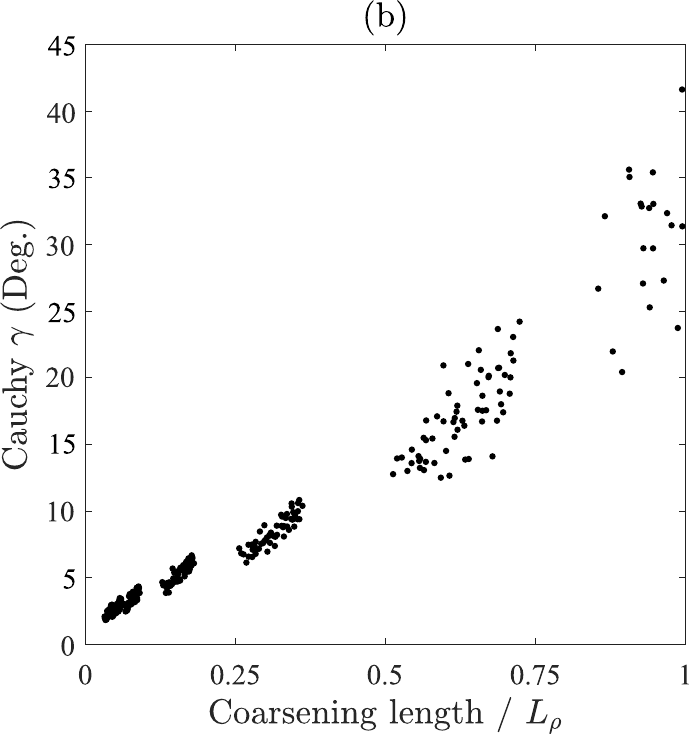}}
    \caption{Trend of distributions with coarse-graining length. The two parameters chosen to describe the shape of the distributions are shown as coarse-graining length is increased: Dirac mass weight at zero fluctuation--the prevalence of zero crossing events--is shown in (a) while the Cauchy parameter--the half-width-at-half-maximum--is shown in (b).}
    \label{fig:distributionfits}
\end{figure}

The trend of coarse-graining leading to distribution broadening can be seen explicitly by fitting the discrete distributions of \cref{fig:Distributions} to an analytical form. The form chosen is:
\begin{equation}
    f(\delta) = P_0 \hat{\delta}(\delta) + (1-P_0) C(\delta\vert\gamma)
\end{equation}
where $C(\delta\vert\gamma)$ is the wrapped Cauchy distribution (\cite{mardiaStatisticsDirectionalData1975}):
\begin{equation}
    C(\delta\vert\gamma) = \frac{1}{2\pi} \frac{\sinh \gamma}{\cosh \gamma - \cos \delta}
\end{equation}
The parameters--the null probability $P_0$ and Cauchy parameter $\gamma$--were obtained by fitting a wrapped Cauchy distribution to the obtained global distributions while omitting the zero fluctuation data point. The results of this fitting process for various coarse-graining lengths are shown in \cref{fig:distributionfits}. In \cref{fig:distributionfits}a, we see the disappearance of the null probability as we move to longer coarse-graining lengths approaching the mean dislocation spacing. This corresponds to a decreased likelihood of single-piercing events as local coarse-graining volumes grow. Similarly, we see in \cref{fig:distributionfits}b the growth of the full-width-at-half-maximum (twice the Cauchy parameter $\gamma$ of the distribution, ranging from only a few degrees for fine scales up to almost a quarter of the unit circle for scales approaching the mean dislocation spacing. This broadening of the distribution indicates the allowance of progressively stronger orientation fluctuations as the coarse-graining process considers larger representative volumes.

\begin{figure*}
    \centering
    \subfloat{\includegraphics[height=2.2in]{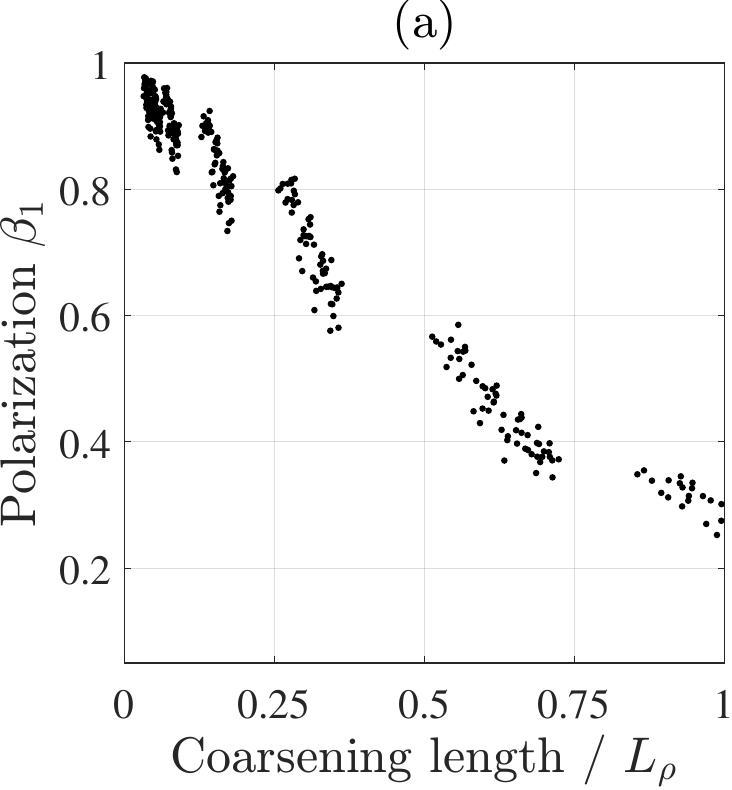}}\hspace{\fill}
    \subfloat{\includegraphics[height=2.2in]{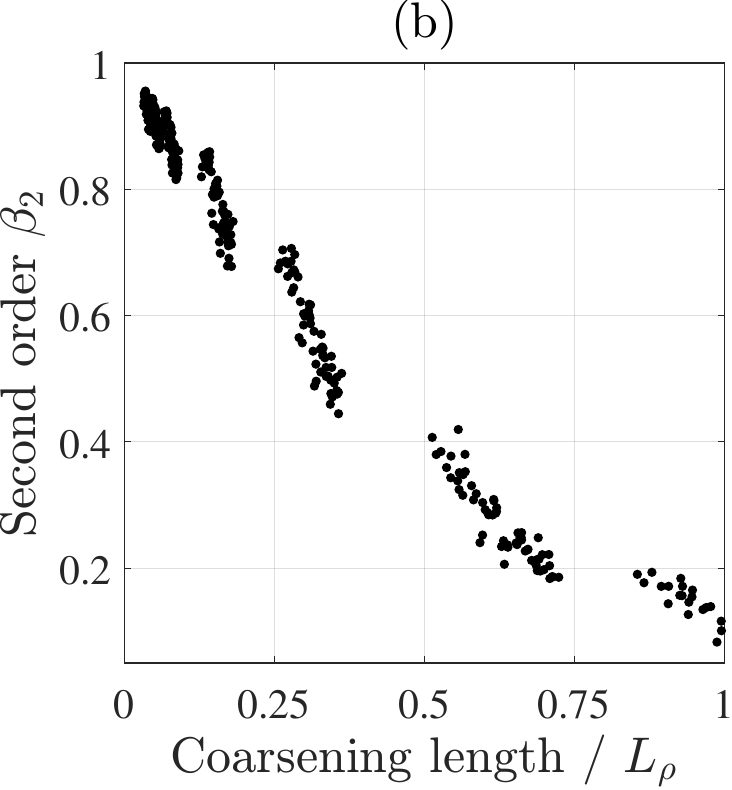}}\hspace{\fill}
    \subfloat{\includegraphics[height=2.2in]{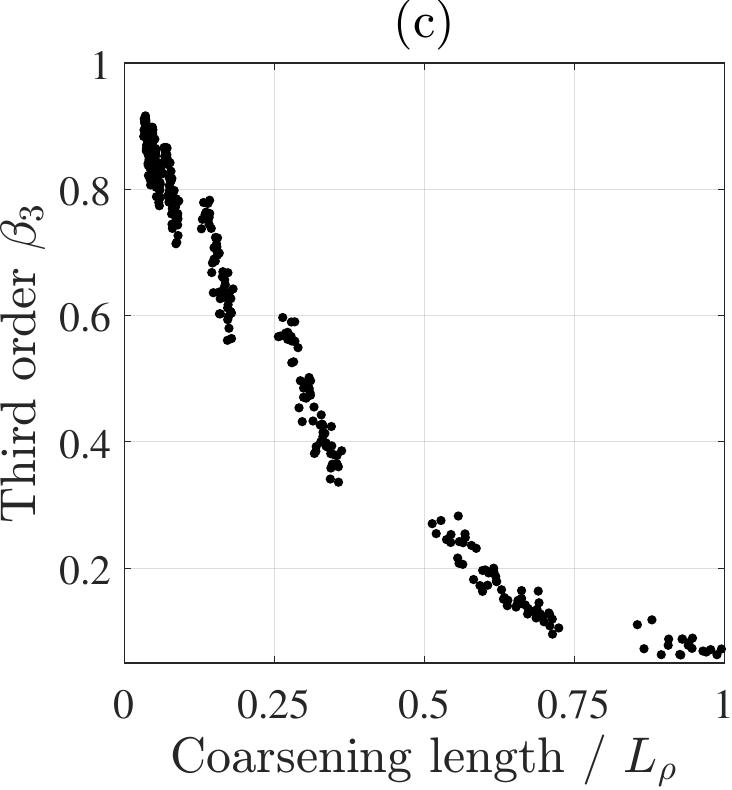}}
    \caption{Characteristic sequence components. The first- (a), second- (b), and third-order (c) characteristic components are plotted against coarse-graining length.}
    \label{fig:characteristicsequence}
\end{figure*}

We may now turn our attention to the characteristic sequences which are the focus of the present work. \Cref{fig:characteristicsequence} shows the first three components in the characteristic sequence for various coarse-graining lengths. Recalling that a perfect Dirac distribution at zero fluctuation has a characteristic sequence everywhere equal to unity, it should not surprise that the fine coarse-graining lengths correspond to near-unit characteristic sequences. As we move to coarser scales, the polarization ($\beta_1$, shown in \cref{fig:characteristicsequence}a) decays. The higher-order components $\beta_2, \ \beta_3$---shown in \cref{fig:characteristicsequence}b-c, respectively---decay more rapidly than the polarization. 

\begin{figure*}
    \centering
    \subfloat{\includegraphics[height=2.5in]{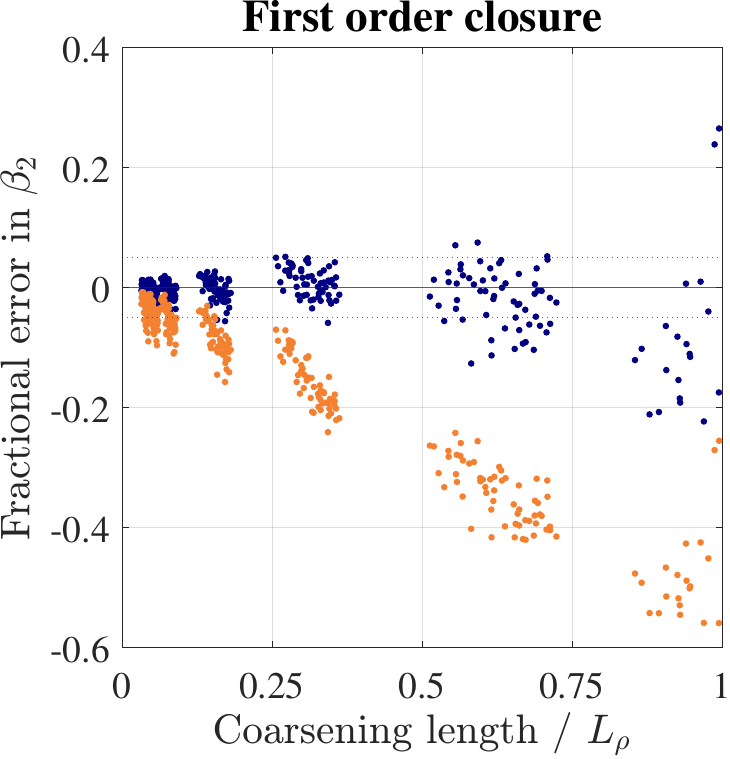}}\hspace{.5in}
    \subfloat{\includegraphics[height=2.5in]{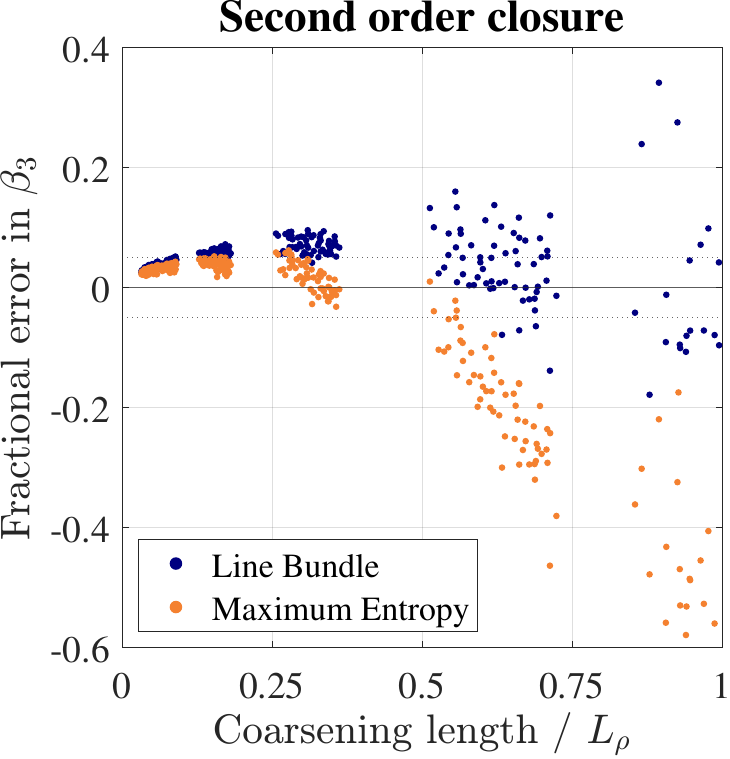}}
    \caption{Fractional closure errors associated with the line bundle and maximum entropy closure forms. These show the fractional error in closing the orientation description at first order (approximating $\beta_2$ as a function of $\beta_1$) and at second order (approximating $\beta_3$ as a function of $\beta_1,\beta_2$) as a function of course graining length.}
    \label{fig:closure}
\end{figure*}

The main question of the present work, however, is how well the closure approximations reproduce the higher-order components. \Cref{fig:closure} shows relative error (approximation/true value) of the line bundle and maximum entropy closure approximations across scales. The line bundle and maximum entropy approximations of the second order component are given by
\begin{align}
    \beta_2^{(\mathrm{LB})} &= \beta_1 ^2 \\
    \beta_2^{(\mathrm{ME})} &= \frac{1}{2}(\beta_1^2+\beta_1^6)
\end{align}
and the third order component by
\begin{align}
    \beta_3^{(\mathrm{LB})} &= \beta_1\beta_2 \\
    \beta_3^{(\mathrm{ME})} &= 2\sqrt{2}\beta_1(\sqrt{1+\beta_2} -3)
\end{align}
\Cref{fig:closure}a shows the relative error in the approximations to the second component of the fluctuation distribution utilized in closing the evolution hierarchy at first order. It is seen that the line bundle closure closely approximates the second order component of the distribution for coarse-graining lengths shorter than half of the mean dislocation spacing. The approximation shows significant variance but still decent agreement with the true second order component for scales up to three-quarters of the mean spacing. The maximum entropy approximation to the second order component, however, performs poorly for all but the finest of coarse-graining lengths, consistently underpredicting the magnitude of the second order component. This is due to the rapid decay of the $\beta_1^6$ term which is simply not present in the ground truth data. {This favoring of the line bundle (wrapped Cauchy-like) over the maximum entropy (assuming von  Mises distribution at first order) first order closure is consistent with the Cauchy rather than von Mises shape of the observed fluctuation distribution functions (cf. \cref{fig:Distributions}b).}

The relative errors for approximations to the third order component of the distribution (second order closure) are shown in \cref{fig:closure}b. At this order, both approximations perform well for coarse-graining lengths shorter than half the mean dislocation spacing. For longer lengths, the line bundle approximation remains a decent description of the third order component, although with the introduction of significant variance. The maximum entropy approximation, on the other hand, quickly begins to deviate from the ground truth for scales above half the dislocation spacing.
\section{Discussion\label{sec:6_disc}}

There are various aspects of the results above that warrant further discussion. First, we will discuss the likelihood that conclusions based on these results can be generalized to a broad range of physical dislocation systems. Secondly, we will attempt to speculate as to why the global distributions found were approximately Cauchy in shape and why the maximum entropy approximation poorly described the shape of the distributions. Following this, we briefly discuss the bearing of the findings on the closure of the evolution hierarchy. Lastly, we discuss a more exotic use case for the fluctuation distributions just reported---a possible implication for continuum treatments of dislocation reaction processes.

\subsection{Applicability of results}\label{applicability-of-results}
The main results of the calculations performed can be summarized as follows. For very small coarse-graining lengths (\(L <\) 200 nm or \(L < L_{\rho}/4\)), the dislocation density is highly polarized (\(\beta_{1} > 0.9\)). This means that the continuum system can be treated analogously to a discrete density field. For intermediate coarse-graining lengths where the polarization deviates substantially from unity (\(1/4 < L/L_\rho < 2/3)\ \)closure of the density hierarchy may be performed appropriately by means of line bundle approximations which assume the characteristic sequence is roughly geometric. For coarse-graining lengths approaching the average dislocation spacing, neither of the candidate closure relations (line bundle or maximum entropy) was appropriate.

For what types of dislocation systems do these findings hold? The systems considered were sparse: the average spacing between dislocations of like slip system in these systems is on the order of \SI{2}{\micro\metre}. As strains progress, the dislocation spacing decreases. In more dense systems, it is unclear whether the angular distributions would be more dependent on the absolute coarse-graining length or the ratio of the coarse-graining length to the average spacing. On the one hand, it would seem that the length relative to the spacing should determine the orientation statistics. However, there also exist absolute length scales associated with physical dislocations. Certain dislocation properties may also affect the orientation statistics, such as reaction lengths or the distance on which oppositely signed dislocations annihilate (often taken to be on the order of 50 nm \cite{Lin2021}). To answer the question of this distinction, one would need discrete simulations with a settled treatment of dislocation reactions and densities on the order of \SI{1000}{\micro\metre^{-2}}. At the moment, this is simply not feasible. In principle, these orientation statistics could be accessible to experiment (cf. dark field x-ray microscopy\cite{simonsDarkfieldXrayMicroscopy2015,poulsenGeometricalopticsFormalismModel2021,borgiSimulationsDislocationContrast2024}) by means of the strain distributions in small crystalline regions, but again, this lies beyond the reaches of the current state-of-the-art. In the meantime, the ambiguity between absolute and relative distances cannot be satisfactorily resolved and we encourage the use of the above results in either sense.

\subsection{The Cauchy shape of the fluctuation distribution}
One of our major findings in the present work concerns the shape of the orientation fluctuation distributions. The maximum entropy assumption results in von Mises-type distributions (generalizations of the Gaussian distribution on a circular domain with higher order moments constrained), while the distributions observed were better represented by (wrapped) Cauchy distributions. This difference in shape resulted in the relatively poor predictive ability for the maximum entropy relations in closing the hierarchy of kinematic variables. The question naturally arises as to why the distribution might be Cauchy instead of maximum entropy in shape. Any answer given to this question will obviously be heuristic and \emph{a posteriori}, but let us attempt one nonetheless. Two possibilities present themselves: 1) the orientation fluctuations are a ratio of Gaussian random variables~\cite{mardiaStatisticsDirectionalData1975} or 2) there are two independent Gaussian processes with different second moments are contributing simultaneously to produce orientation fluctuations~\cite{baileyEmergenceWrappedCauchy2021}. We will focus on the former. If the likelihood of a fluctuation angle is related to its ratio to a representative angular scale which is also a random variable, the Cauchy shape may be produced. The natural candidate for this is curvature, which determines the spatial strength of changes in line orientation. Considering a segment of uniform curvature passing through a volume (a circular arc). The orientations would be uniformly distributed on an interval \(\delta \in \left\lbrack - |\kappa l|/2,|\kappa l|/2 \right\rbrack\) where \(\kappa\) is the curvature of the arc and \(l\) is the length of the arc; the term \(|\kappa l|\) is properly related to the local curvature density \(q\). If the likelihood of a fluctuation is related to the weight it contributes to a point relative to the curvature density of the segment and both are random variables, a Cauchy shape could result. Another line of reasoning might proceed from Gibbs-type energy penalization of perturbations of dislocation lines via their change to the arc-length of the dislocation line, which would also depend inversely on curvature.

Nonetheless, any additional information regarding the shape of the distribution cuts against the maximum entropy problem approach. The maximum entropy approach assumes that the only knowledge we have of the system, qualitative or quantitative, is the first \(n\) moments of the distribution. Prior knowledge of the orientation fluctuation as a rational process would then have to be taken into account as an additional constraint in the entropy maximization process. Thus, the strong leaning of the distributions towards a Cauchy-like shape points to at least a qualitative `hidden variable'; that is, something we \emph{ought} to know about the orientation fluctuation process before simply leaping into the entropy maximization operation. Even at the large coarse-graining lengths the maximum entropy approach was designed to treat, we see it break down. This is presumably due to some important hidden information regarding the process that is not captured simply by the first few moments. Especially notable was the poor performance of the first order maximum entropy closure due to a quickly decaying term which was not present in the data. I would guess that this poor agreement was what led to the generally poor performance of the first order closure in the initial dynamical benchmarking study of the maximum entropy closure~\cite{Monavari2016}.

\subsection{Implications for the evolution equations}
The different length scale regimes noted above warrant different treatments of the dislocation dynamics. Regardless of which regime is being considered, all of these treatments amount to a closure of the density hierarchy at some fixed order. However, inspection of the evolution equations \cref{eq:11_reducedEvolution} reveals that the auxiliary curvature tensor hierarchy also needs to be similarly closed. In the case that the curvature is homogeneous in the orientation coordinate, the auxiliary curvature tensors can be expressed as:
\begin{subequations}
\begin{align}
\mathbf{Q}^{(n)}(\boldsymbol{r}) &= \kappa\left( \boldsymbol{r} \right)\rho\left( \boldsymbol{r} \right)\int_{S^{1}}^{}{f_{\boldsymbol{r}}(\varphi)\boldsymbol{p}^{\otimes2}(\varphi)\boldsymbol{l}^{\otimes n - 2}(\varphi)d\varphi}\\    
&= q\left( \boldsymbol{r} \right)\ \text{sym}\left\lbrack \widehat{\boldsymbol{n}}\times\left\langle \boldsymbol{l}^{\otimes n} \right\rangle_{\boldsymbol{r}} \times \widehat{\boldsymbol{n}} \right\rbrack
\end{align}
\end{subequations}
Still, this requires fixing the curvature \(\kappa\boldsymbol{(r)}\). It has been shown~\cite{Sedlacek2010,Hochrainer2010,zhangContinuumDislocationDynamics2025} that in the purely polarized state, the curvature is given by:
\begin{equation}
\kappa\left( \boldsymbol{r} \right) \coloneqq {\overline{\boldsymbol{l}}}_{\boldsymbol{r}}\cdot\nabla{\overline{\varphi}}_{\boldsymbol{r}}
\end{equation}
or in a coordinate-independent form
\begin{equation}
\kappa\left( \boldsymbol{r} \right)=\left( \hat{\boldsymbol{n}}\times\nabla \right)\cdot{\overline{\boldsymbol{l}}}_{\boldsymbol{r}}
\end{equation}
This can be seen by expanding the time derivative of the vector magnitude of the first order evolution equation {[}\cref{eq:11_reducedEvolution}{]} and comparing it to the zeroth order equation. This expression also holds for systems where the polarization gradients are small.

Although we have seen that there are highly polarized regimes of dislocation dynamics (where \(L < 200\) nm), the more significant range is that where the polarization is non-negligible but the line bundle closure is still valid (\(L < 2L_{\rho}/3\)). In this case, it would be worthwhile to express the curvature in terms of its completely polarized contribution {[}Eq. (21){]} as well as some additional curvature source tied to the `statistically stored' portion of the density (\(\epsilon(\boldsymbol{r)} = 1 - \beta_{1}(\boldsymbol{r})\)). While the closure of the density hierarchy at small scales has been demonstrated in the present work, a definitive linking of the discrete scale to the macroscopic case (where curvature must be treated as a dynamic quantity) would involve fixing the curvature in terms of density-based quantities (as is done at large coarse-graining lengths in~\cite{zhangContinuumDislocationDynamics2025}).

Assuming this discrete expression for the curvature, one finds that in systems with high or constant polarization, the discrete curl-type transport equation is sufficient to describe the system. This treatment we have been referring to as vector density dislocation dynamics\cite{Anderson2022}. Nonetheless, the line bundle assumption still has implications even for the curl transport equation. When one moves to a consideration of kinetics, the velocity is dependent on the line direction by means of the Peach Koehler force:
\begin{equation}
\boldsymbol{v}=M\left( \boldsymbol{b} \cdot \boldsymbol{\sigma}\left( \boldsymbol{r} \right) \right)\times \boldsymbol{l}
\end{equation}
Upon averaging the transport equation, the second-order alignment tensor thus appears:
\begin{subequations}
\begin{align}
    \partial_{t}\boldsymbol{\rho}&=\nabla\times\left\langle \boldsymbol{v}\times\rho\boldsymbol{l} \right\rangle\\
    &= M\nabla \times \left\langle \left( \left( \boldsymbol{b} \cdot \boldsymbol{\sigma} \right)\times\boldsymbol{l} \right)\times \boldsymbol{l} \right\rangle\\
    &=M\ \nabla \times \left\langle \boldsymbol{b}\cdot \boldsymbol{\sigma} \cdot \left( \boldsymbol{l}^{\otimes2}-\mathbb{ I} \right)\rho \right\rangle
\end{align}
\end{subequations}
Averages of this form are the motivation for the more general form of the line bundle assumption utilized in the appendix. This version allows a single factor of \(\boldsymbol{l}\) to be factored out of such averages, resulting in a kinetic equation expressible in terms of the vector density and a suitable correlation function. For a more detailed treatment of this question of kinetics, we refer the reader to a previous work\cite{andersonDislocationCorrelationsContinuum2024}.

Beyond the quasi-discrete regime of high polarization, the present findings anticipate new treatments of continuum dislocation dynamics at intermediate length scales. Closing the evolution equations at second order while utilizing the discrete curvature expression is a new possibility. Corrections to the discrete evolution equation based on weak gradients in the polarization field may also be considered. We hope that the present work inspires new hybrid approaches blending the quasi-discrete vector density transport and the hopelessly general higher-order density transport. It is our opinion that progress toward an understanding of fundamental deformation phenomena at small scales---such as fatigue, the onset of dislocation patterning, and even crack dynamics---will come from hybrid approaches in the intermediate regime where the line bundle assumption holds with non-trivial polarizations. In this spirit, we next turn our attention to how the present results would inform the physics of dislocation interactions in such an intermediate regime.

\subsection{Implications for continuous dislocation reactions}
{Dislocation reactions result in interchange of density between slip systems, resulting in the coupling of dislocation evolution equations in a manner analogous to reaction-diffusion methods:} 
\begin{align}
\dot{\boldsymbol{\rho}}^{[\alpha]}_j=&\dot{\boldsymbol{\rho}}^{[\alpha]}_\text{glide}  +\dot{\boldsymbol{\rho}}^{[\alpha],\text{cs}} +\dot{\boldsymbol{\rho}}^{[\alpha],\text{gj}} -\dot{\boldsymbol{\rho}}^{[\alpha],\text{sj}}\\
(\dot{\boldsymbol{\rho}}^{[\alpha],\text{cs}})_j =& \rho^{[\alpha']}_j\dot{r}^{[\alpha',cs]}\left(\bar{\psi}^{[\alpha']},\sigma_{ij}(\boldsymbol{r})\right) \nonumber \\&- \rho^{[\alpha]}_j\dot{r}^{[\alpha,cs]}\left(\bar{\psi}^{[\alpha]},\sigma_{ij}(\boldsymbol{r})\right)\\
(\dot{\boldsymbol{\rho}}^{[\alpha],\text{gj}})_k =& \sum_{\beta,\gamma} \rho^{[\beta]}_i\rho^{[\gamma]}_j \dot{r}^{[\beta,\gamma\to\alpha, \text{gj}]}_{ij}(\bar\psi^{[\beta]},\bar\psi^{[\gamma]}) \nonumber \\
&- \rho^{[\alpha]}_i\sum_\beta\dot r^{[\alpha,\beta\to \gamma, \text{gj}]}_{ij}(\bar\psi^{[\alpha]},\bar\psi^{[\beta]})\\
(\dot{\boldsymbol{\rho}}^{[\alpha],\text{sj}})_k =&  \rho^{[\alpha]}_i\sum_{\beta}\dot \rho_j^{[\beta]}r^{[\alpha,\beta\to s]}_{ijk}(\bar\psi^{[\alpha]},\bar\psi^{[\beta]})
\end{align}
{
where $\dot{\boldsymbol{\rho}}_\text{slip}$ is the density evolution due to dislocation glide that we have been thus far discussing, and $\dot{\boldsymbol{\rho}}^{[\alpha],\text{cs}} ,\dot{\boldsymbol{\rho}}^{[\alpha],\text{gj}},\dot{\boldsymbol{\rho}}^{[\alpha],\text{sj}}$ are the net production (positive) or loss (negative) of dislocation density due to various reactions (cross-slip, glissile junction, and sessile junction). The rates of these reactions $\dot r$ generally vary based on the type of reaction and are dependent on the orientation of the reacting systems' density vectors with respect to the slip plane dihedral $(\bar{\psi})$. In the case of cross slip, each system $\alpha$ has only one cross-slip system $\alpha'$, and the rate is additionally dependent on the local stress state. The interested reader is referred to \cite{Vivekanandan2021} for a more detailed discussion of these rates.}

\begin{figure*}
    \centering
    \includegraphics[width=0.8\linewidth]{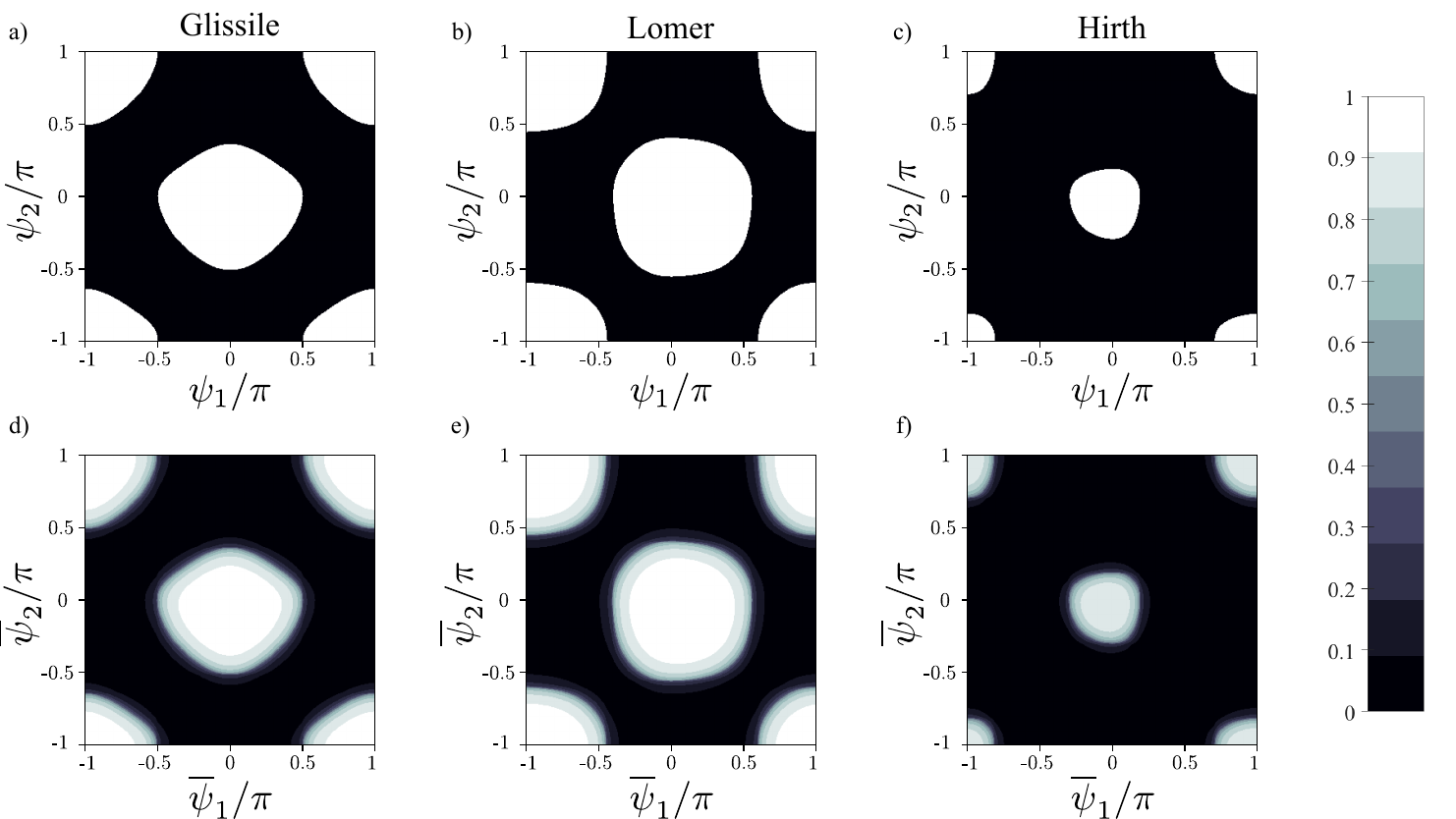}
    \caption{Discrete and Continuum reaction maps. The pairs of angles (measured with respect to the dihedral direction) which result in the three junction types---glissile junction, Lomer lock, Hirth lock---for discrete dislocation segments are shown in (a-c), respectively. The corresponding maps for continuum dislocation density vectors (d-f) give the likelihood that discrete segments in the collection fall within the reaction region. The orientation fluctuation distribution functions used to obtain the continuum reaction maps were calculated using a 97 nm coarse-graining length.}
    \label{fig:reactionmaps}
\end{figure*}

In discrete dislocation systems, there are several modes by which dislocations might react with each other to form junction segments. This process begins with a crossed state, where two dislocations on separate slip planes intersect. These two dislocations are described by their respective angles \(\psi_{1},\psi_{2}\) with respect to the dihedral formed by the two glide planes. In certain regions of the two-dimensional orientation space, this crossed state is unstable: the formation of a junction segment along the dihedral would lower the elastic energy of the system. Thus, the reaction is considered to be a deterministic process:  if the energy criterion is satisfied, the crossed state always forms a junction; if not, the dislocations glide past each other and form a jog. The deterministic reaction maps (using the energy criterion of~\cite{Lin2020,madecNatureAttractiveDislocation2002}) are shown in \cref{fig:reactionmaps}a-c: white denotes the formation of a junction while black denotes jog formation. We consider these maps to be a function taking unit value in the reacting region and null in the complement:
\begin{equation}
\dot r=1_{\Delta E < 0}\left( \psi_{1},\psi_{2} \right) = 
\begin{dcases}
1 & \text{for junction formation},  \\
0 & \text{otherwise}.
\end{dcases} 
\end{equation}
In the continuum system, we do not have access to the direction of each individual dislocation at a given point. We do, however, have access to the average direction of dislocations at a given point \({\overline{\psi}}_{1},{\overline{\psi}}_{2}\). Additionally, we know the likelihood of fluctuations from that direction in an average sense. {Additionally, the relatively small size of these angular fluctuations allows the preservation of analogous behavior to the discrete system. As a result, the fluctuation information can be used to determine the likelihood of a dislocation reaction for a given pair of average directions:}
\begin{widetext}
\begin{equation}
\dot{r} =P_{j}^{(L)}\left( {\overline{\psi}}_{1},{\overline{\psi}}_{2} \right) = \int_{0}^{2\pi}{d\delta_{1}\int_{0}^{2\pi}{d\delta_{2}}f^{(L)}\left( \delta_{1} \right)f^{(L)}\left( \delta_{2} \right)1_{\Delta E < 0}\left( \overline{\psi}_{1} + \delta_{1},{\overline{\psi}}_{2} + \delta_{2} \right)}.
\end{equation}
\end{widetext}
This is simply the convolution of the discrete reaction map with the joint fluctuation distribution (the two fluctuation processes are independent\footnote{Similar analysis to section 4 was performed to calculate the joint distributions for dislocations within a cutoff radius; no deviation from the independent joint distribution was found even for small cutoff radii.}). The resulting continuous reaction maps are shown in \cref{fig:reactionmaps}d-f.

One of the central results of this work was the observation that local fluctuations in dislocation orientation were of finite strength and that
their distribution could be measured. The implication is that at any continuum scale, orientation-dependent processes must consider this uncertainty in the precise directions of the underlying dislocations. This holds not only for deterministic processes such as the reactions just considered, but also for probabilistic processes such as cross-slip~\cite{vivekanandanDataDrivenApproach2023}. This example serves to show that the implications of orientation fluctuations reach beyond the closure of the evolution hierarchy.

\section{Conclusion}
In the present work, we have put forward a treatment of the average strength of orientation fluctuations in a spatially coarse-grained continuum treatment of dislocations. A key feature of this analysis was the fluctuation distribution function which describes the distribution of small deviations from the local average dislocation orientation. These fluctuations are closely related to the coarse-graining length, as it controls the size of the region wherein dislocations are treated collectively. In addition to the local fluctuation distribution, the total volume average of this distribution---the global fluctuation distribution---was put forward to explicitly analyze the dependence of local orientations on coarse-graining length in an average sense.

The importance of the fluctuation distributions governing deviations from the local average orientation on the transport was discussed in the context of the higher-order treatment of the evolution of dislocations distributed across orientation~\cite{Hochrainer2015}. A key concept in this treatment is the characteristic sequence of the distribution (its Fourier coefficients). A reduced form of the higher-order transport is given by closing this sequence at low order. Two such closure approximations were considered. First, the line bundle approximation---novel to the present work---was derived from assuming neighboring entries in the characteristic sequence are geometric, a trait we refer to as Cauchy-like. For comparison, we also evaluated the maximum entropy approximation due to Monavari et al.~\cite{Monavari2016}. It was also shown that the validity of the closure approximations at a given coarse-graining length could be evaluated from the global fluctuation distribution.

Measurements of the global fluctuation distribution were performed using discrete dislocation dynamics simulations. These simulations were analyzed at various coarse-graining lengths. The resulting distributions were approximately Cauchy in shape, having an additional spike at zero fluctuation due to the discretization of the dislocation segments. At small coarse-graining lengths, this distribution was sharply concentrated at zero fluctuations, with the distribution broadening at larger coarse-graining lengths. An analysis of the characteristic sequence of the measured distributions found that very short coarse-graining lengths (\(L < 200\ nm\ \)or 1/4 the average dislocation spacing) had trivial polarizations, implying closure at first order may be sufficient. For longer length scales (1/4-2/3 the average dislocation spacing), the second and third order line bundle approximations were seen to be an adequate description of the respective components of the characteristic sequence. {The maximum entropy approximation did not well describe the second order component for any regime of coarse-graining length among those considered and only reasonably described the third order component for smallest coarse-graining lengths considered. To our surprise, the maximum entropy approximation actually performed worse as more dislocations were included in a local collection, with the worst observed agreement at coarse-graining lengths approaching and exceeding the mean dislocation spacing. For this approximation to be accurate at large coarse-graining lengths, some mechanism must present itself at scales greater than the typical simulation volume of discrete dislocation dynamics simulations that would reverse this trend.}

The future of continuum dislocation dynamics modeling is likely to be found in the intermediate length scale regimes discussed in the present work. The mesoscale approach, towards which this represents the first steps, will be necessary to describe dislocation dynamics at intermediate scales where the perfectly polarized transport (vector density dislocation dynamics) needs correction, but the full higher-order transport is redundantly verbose. While global orientation fluctuation distributions presented here inform a closure of the density hierarchy at these small scales, further investigation into the curvature distribution at these small scales is needed to form a complete picture of dislocation dynamics at these intermediate scales. Additionally, orientation-dependent processes (e.g., cross-slip and dislocation reactions) ought to be informed by these fluctuation distributions as well at these intermediate length scales. We see this work as a first step in bridging the vector density and higher-order dislocation transport frameworks, showing that the former is appropriate at short length scales and the latter at long length scales. Rather than two separate theories, they exist at different ends of a spectrum; insights from both can inform our treatment of continuum dislocation dynamics at intermediate regimes.

\section*{Data availability statement}
The data that support the findings of this article, as well as the scripts that reproduce the figures, are openly available \cite{dataset}.

\section*{Acknowledgment}
This work was supported by the US Department of Energy, Office of Science, Division of Materials Sciences and Engineering, through award number DE-SC0017718 at Purdue University, and by the US Department of Energy, Office of Science, Office of Fusion Energy Sciences via award number DE-SC0024585 at Purdue University.

\appendix

\section{Equivalence of moment/Fourier representations of the shape of the distribution}
The equivalence of the alignment tensor series and the characteristic sequence can be seen by simply expanding the expression of the unit tangent in the local coordinate system defined by the average direction {[}cf. \cref{eq:4avorient}{]}. In this form, it can be clearly seen that
\begin{subequations}
\begin{align}
    \boldsymbol{l}(\delta) &= \cos\delta\overline{\boldsymbol{l}} + \sin\delta\overline{\boldsymbol{p}}\\
    &=\boldsymbol{L}e^{i\delta} + \boldsymbol{L}^{*}e^{- i\delta}\\
    \boldsymbol{L} &\coloneqq \frac{1}{2}\left( \overline{\boldsymbol{l}} - i\overline{\boldsymbol{p}} \right)
\end{align}
\end{subequations}
The tensor powers \(\boldsymbol{l}^{\boldsymbol{\otimes}n}(\delta)\) can then be expanded by means of the binomial theorem for either the complex or trigonometric form. In the case of the former, we obtain:
\begin{equation}
\boldsymbol{l}^{\boldsymbol{\otimes}n}(\delta) = \sum_{k = 0}^{n}{{n\choose k}\overline{\boldsymbol{L}^{\boldsymbol{\otimes}k}\boldsymbol{L}^{\boldsymbol{* \otimes}n - k}}e^{(2k - n)i\delta}}
\end{equation}
Upon taking the average we obtain 
\begin{equation}
\left\langle \boldsymbol{l}^{\boldsymbol{\otimes}n}(\delta) \right\rangle_{\boldsymbol{r}}^{(L)} = \sum_{k = 0}^{n}{{n\choose k}\overline{\boldsymbol{L}^{\boldsymbol{\otimes}k}\boldsymbol{L}^{\boldsymbol{* \otimes}n - k}}\beta_{2k - n}}
\end{equation}
We have used now twice the symmetrizing operator, which averages over
all permutations of indices:
\begin{equation}
{\overline{A}}_{i_{1}i_{2}\ldots i_{n}} = \text{sym}\left\lbrack A_{i_{1}i_{2}\ldots i_{n}}\  \right\rbrack \coloneqq \frac{1}{n!}\sum_{\pi \in \Pi_{n}}^{}A_{\pi_{1}\pi_{2}\ldots\pi_{n}}\end{equation}
Equivalently, we may expand the tensor power in terms of the
trigonometric functions: 
\begin{equation}
\boldsymbol{l}^{\boldsymbol{\otimes}n}(\delta) = \sum_{k = 0}^{n}{{n\choose k}\overline{{\overline{\boldsymbol{p}}}^{\boldsymbol{\otimes}k}{\overline{\boldsymbol{l}}}^{\boldsymbol{\otimes}n - k}}\cos^{n - k}\delta\sin^{k}\delta}.
\end{equation}
There is clearly a relation between these in terms of the expansion of
powers of the complex forms of sine and cosine.

From these expressions for the tensor powers, one sees their
average---the alignment tensors---can be expressed simply as:
\begin{subequations}
\begin{align}
    \left\langle \boldsymbol{l}^{\boldsymbol{\otimes}n}(\delta) \right\rangle &= \sum_{k = 0}^{n}{{n \choose k} \overline{\boldsymbol{L}^{\boldsymbol{\otimes}k}\boldsymbol{L}^{\boldsymbol{* \otimes}n - k}}\beta_{2k - n}}\\
    &= \sum_{k = 0}^{n}{{n\choose k}\overline{{\overline{\boldsymbol{p}}}^{\boldsymbol{\otimes}k}{\overline{\boldsymbol{l}}}^{\boldsymbol{\otimes}n - k}}M^{'}(n,k)}\\
    M^{'}(n,k) &= \left\langle \cos^{n - k}\delta\sin^{k}\delta \right\rangle\\
&=  \begin{dcases}
\sum_{i = 0}^{j}{{j\choose i}( - 1)^{i}M_{n - 2i}} & k \text{\ even} \\
\sum_{i = 0}^{\left\lfloor \frac{n - 1}{2} \right\rfloor - j}{{\left\lfloor \frac{n - 1}{2} \right\rfloor - j}\choose i}( - 1)^{i}{\widetilde{M}}_{n - 2i} & k \text{\ odd}
\end{dcases}.
\end{align}
\end{subequations}
For what we hope are obvious reasons, we choose to work in terms of the
characteristic sequence \(\beta_{k}\).

\section{Derivation of a general class of line bundle approximations}
We would like to consider a general class of averages having the form: 
\begin{equation}
\left\langle A(\delta)\boldsymbol{l}^{\otimes n}(\delta) \right\rangle \label{eq:MainAv}
\end{equation}
where \(A(\delta)\) is some general function of the orientation
fluctuations. Under what conditions can we remove a factor of
\(\boldsymbol{l}\) from the average? That is, when do we have
\begin{equation}
\left\langle A(\delta)\boldsymbol{l}^{\otimes n}(\delta) \right\rangle\boldsymbol{\approx}\left\langle A(\delta)\ \text{sym}\left\lbrack \overline{\boldsymbol{l}} \otimes \boldsymbol{l}^{\otimes n - 1}(\delta) \right\rbrack \right\rangle \label{eq:LBdef}
\end{equation}
We can see that the closure of the alignment tensor hierarchy is a special case of this where \(A\) is a constant unit function.

The only variable available to us with which to define this approximation is the fluctuation distribution \(f(\delta)\) and its characteristic sequence \(\beta_{n}\). As a result, this approximation will be better formulated in Fourier space. The average of interest [\cref{eq:MainAv}] is, in Fourier notation:
\begin{equation}
\mathfrak{F}\left\{ f(\delta)l^{\otimes n}(\delta)A(\delta) \right\}_{0} = \sum_{k = - \infty}^{\infty}{\gamma_{k}^{(n)}{\hat{A}}_{- k}}
\end{equation}

where we have denoted the Fourier transform by \(\mathcal{F}\) or by \(\hat{\cdot}\) and utilized the convolution theorem. Two key
Fourier series are denoted \(\gamma_{k}^{(n)}\) and \(\lambda_{k}^{(n)}\) and are given by 
\begin{align}
\gamma_{k}^{(n)}&\coloneqq\mathfrak{F}\left\{ f(\delta)l^{\otimes n}(\delta) \right\}_{k}\\
&= \sum_{m = - \infty}^{\infty}{\lambda_{m}^{(n)}\beta_{k - m}}\label{eq:gamma_def}\\
\lambda_{k}^{(n)}&\coloneqq\mathfrak{F}\left\{ l^{\otimes n}(\delta) \right\}_{k}
\end{align}
It should be noted that both \(\gamma_{k}^{(n)}\) and \(\lambda_{k}^{(n)}\) are fully symmetric \(n\)-th rank tensors. Thus the goal of our analysis should be to show a circumstance whereby:
\begin{equation}
\gamma_{k}^{(n)} \approx \overline{\boldsymbol{l}} \otimes \gamma_{k}^{(1)}
\end{equation}

Before proceeding, we should consider the form of \(\lambda_{k}^{(n)}\), as it is fixed by the definition of \(l\) in \cref{eq:orientationDef,eq:orientationExponential}. Especially, the convenient exponential structure of \cref{eq:orientationExponential} is especially suited to defining the Fourier series \(\lambda_{k}^{(n)}\). The series first few \(n\) are as follows:
\begin{subequations}
    \begin{align}
\lambda_{k}^{(0)}&=\mathfrak{F}\left\{ 1 \right\}_{k} = \delta_{k}\\
\lambda^{(1)}&=\mathfrak{F}\left\{ l(\delta) \right\}_{k}\\
&= L\delta_{k - 1} + L^{*}\delta_{k + 1}\\
\lambda^{(2)}&=\mathfrak{ F}\left\{ l(\delta) \otimes l(\delta) \right\}_{k}\\
&= LL\delta_{k - 2} + \left( LL^{*} + L^{*}L \right)\delta_{k} + L^{*}L^{*}\delta_{k + 2}\\
&= L\lambda_{k - 1}^{(1)} + L^{*}\lambda_{k + 1}^{(1)}\end{align}
\end{subequations}
In fact, due to the repeated multiplication by the complex exponentials
contained in the definition of \(l\), one can see that in general
\begin{equation}
\lambda_{k}^{(n)} = \text{sym}\left\lbrack L\lambda_{k - 1}^{(n - 1)} + L^{*}\lambda_{k + 1\ }^{(n - 1)} \right\rbrack\end{equation}
Although we will omit the symmetric notation for brevity.

Now that we have a recursion relation for \(\lambda_{k}^{(n)}\), we can use this in the definition of the \(\gamma_{k}^{(n)}\) series [\cref{eq:gamma_def}]:
\begin{subequations}
\begin{align}
\gamma_{k}^{(n)} &= \sum_{m = - \infty}^{\infty}{\left( L\lambda_{m - 1}^{(n - 1)} + L^{*}\lambda_{m + 1\ }^{(n - 1)} \right)\beta_{k - m}}\\
&= \sum_{m = - \infty}^{\infty}{\left( L\beta_{k - m - 1} + L^{*}\beta_{k - m + 1} \right)\lambda_{m}^{(n - 1)}}\label{eq:19_gamma_int1}
\end{align}
\end{subequations}
where we have simply reindexed the summation to transfer the shift to
the characteristic sequence. It now becomes clear that our constraint on
the characteristic function should be in the form of a recursion
relation for \(\beta_{n}\). Let us denote this relation by the
ladder operators \(\eta_{k}^{\pm}\), defined such that:
\begin{equation}
\eta_{k}^{\pm}\beta_{k} = \beta_{k \pm 1}
\end{equation}
\Cref{eq:19_gamma_int1} above then becomes in:
\begin{equation}
\gamma_{k}^{(n)} = \sum_{m = - \infty}^{\infty}{\left( L\eta_{k - m}^{-} + L^{*}\eta_{k - m}^{+} \right)\beta_{k - m}\lambda_{m}^{(n - 1)}}\label{eq:19_gamma_int2}
\end{equation}
Note that if the quantity in parentheses can be removed from the summation, we find that the remaining summation is equivalent to \(\gamma_{k}^{(n - 1)}\) by definition. We will propose one form of the ladder operators that allows this to be performed.

By malice of forethought, we propose the following ladder operators:
\begin{equation}
\eta_{k}^{\pm} = \beta\delta_{k} + \beta^{\pm 1}\theta_{k}^{+} + \beta^{\mp 1}\theta_{k}^{-}
\end{equation}
where \(\theta_{k}^{\pm}\) is unity for strictly positive (negative) \(k\) and zero otherwise. These would be the raising and lowering operators that result if the characteristic function is a geometric series:
\begin{equation}
{\widetilde{\beta}}_{k} = \beta^{|k|}
\end{equation}
where \(\beta < 1\ \)(i.e. \(f\) is Cauchy). The approximation is not that \(\beta_{k} \approx {\widetilde{\beta}}_{k}\) for all \(k\),
but rather that
\begin{equation}
\eta_{k}^{\pm}\beta_{k} \approx \beta_{k \pm 1}.
\end{equation}

The complex unit vectors in \cref{eq:19_gamma_int2} can be expanded to find that we are interested in the sum and difference of the ladder operators:
\begin{equation}
L\eta_{k}^{-} + L^{*}\eta_{k}^{+} = {\overline{\eta}}_{k}\overline{l} + {\widetilde{\eta}}_{k}i\overline{p}
\end{equation}
With the form of the ladder operators proposed above, the sum and difference are given by:
\begin{subequations}
\begin{align}
{\overline{\eta}}_{k} &= \frac{1}{2}\left( \eta_{k}^{+} + \eta_{k}^{-} \right) = \overline{\beta} + \widetilde{\beta}\delta_{k}    \\
{\widetilde{\eta}}_{k} &= \frac{1}{2}\left( \eta_{k}^{+} - \eta_{k}^{-} \right) = \widetilde{\beta}\left( \theta_{k}^{+} - \theta_{k}^{-} \right)
\end{align}
\end{subequations}
where 
\begin{subequations}
\begin{align}
    \overline{\beta} &= \frac{1}{2}\left( \beta + \beta^{- 1} \right)\\
    \widetilde{\beta} &= \frac{1}{2}\left( \beta - \beta^{- 1} \right)
\end{align}
\end{subequations}
We note that in terms of an expansion in terms of \(\beta = 1 - \epsilon\), these are
\begin{subequations}
\begin{align}
    \overline{\beta} &= 1 + \frac{\epsilon^{2} + \epsilon^{3} + \ldots}{2}\\
    \widetilde{\beta} &= - \left( \epsilon + \frac{\epsilon^{2} + \epsilon^{3} + \ldots}{2} \right).
\end{align}
\end{subequations}
Returning to \cref{eq:19_gamma_int2}, then, we obtain an expression for \(\gamma_{k}^{(n)}\) in terms of the next lower order:
\begin{align}
\gamma_{k}^{(n)} &= \overline{\beta}\ \text{sym}\left\lbrack \overline{l}\gamma_{k}^{(n - 1)} \right\rbrack - \widetilde{\beta}\ \text{sym}\left\lbrack \overline{\boldsymbol{l}}\lambda_{k}^{(n - 1)} - i\overline{\boldsymbol{p}}{\widetilde{\gamma}}_{k}^{(n - 1)} \right\rbrack\\
{\widetilde{\gamma}}_{k}^{(n - 1)} &\coloneqq \sum_{m = - \infty}^{\infty}{\left( \theta_{k - m}^{-} - \theta_{k - m}^{+} \right)\beta_{k - m}\lambda_{m}^{(n - 1)}}\\
&= \sum_{m = 1}^{\infty}{\beta_{m}\left( \lambda_{k + m}^{(n - 1)} - \lambda_{k - m}^{(n - 1)} \right)}
\end{align}
We do see that the leading order term is the desired result. However,
there is a first order correction in terms of the antisymmetric portion
of the average \(\widetilde{\gamma}.\) Thus we can only neglect these
corrections if the polarization is approximately unity even to first
order.

The \(\gamma_{k}^{(n)}\ \)expression can be evaluated at \(k = 0\ \)to
give the alignment tensor closure relation. For \(n = 2,3\) this is as
follows:
\begin{align}
    \gamma_{0}^{(2)} &= \frac{1}{2}\left\lbrack \left( 1 + \beta_{1}^{2} \right)\overline{\boldsymbol{ll}} + \left( 1 - \beta_{1}^{2} \right)\overline{\boldsymbol{pp}} \right\rbrack\\
    \gamma_{0}^{(3)} &= \frac{\beta_{1}}{4}\left\lbrack \left( 3 + \beta_{2} \right)\overline{\boldsymbol{lll}} + \left( 2\beta_{1}^{- 2} + 1 - \beta_{2} \right)\overline{\boldsymbol{lpp}} \right\rbrack.
\end{align}

\bibliography{AngStatsBib}

@article{Acharya2006,
  title = {Size Effects and Idealized Dislocation Microstructure at Small Scales: {{Predictions}} of a {{Phenomenological}} Model of {{Mesoscopic Field Dislocation Mechanics}}: {{Part I}}},
  author = {Acharya, Amit and Roy, Anish},
  year = {2006},
  month = aug,
  journal = {Journal of the Mechanics and Physics of Solids},
  volume = {54},
  number = {8},
  pages = {1687--1710},
  issn = {00225096},
  doi = {10.1016/j.jmps.2006.01.009}
}

@article{Anderson2021,
  title = {On the Three-Dimensional Spatial Correlations of Curved Dislocation Systems},
  author = {Anderson, Joseph Pierre and {El-Azab}, Anter},
  year = {2021},
  month = mar,
  journal = {Materials Theory},
  volume = {5},
  number = {1},
  pages = {1--34},
  publisher = {SpringerOpen},
  issn = {2509-8012},
  doi = {10.1186/S41313-020-00026-W},
  urldate = {2021-10-14}
}

@article{Anderson2022,
  title = {Situating the {{Vector Density Approach Among Contemporary Continuum Theories}} of {{Dislocation Dynamics}}},
  author = {Anderson, Joseph Pierre and Vivekanandan, Vignesh and Lin, Peng and Starkey, Kyle and Pachaury, Yash and {El-Azab}, Anter},
  year = {2022},
  month = jan,
  journal = {Journal of Engineering Materials and Technology},
  volume = {144},
  number = {1},
  publisher = {American Society of Mechanical Engineers Digital Collection},
  issn = {0094-4289},
  doi = {10.1115/1.4052066},
  urldate = {2022-03-18}
}

@article{andersonDislocationCorrelationsContinuum2024,
  title = {Dislocation Correlations and the Continuum Dynamics of the Weak Line Bundle Ensemble},
  author = {Anderson, Joseph Pierre and {El-Azab}, Anter},
  year = {2024},
  journal = {Physical Review B},
  volume = {109},
  number = {17},
  doi = {10.1103/PhysRevB.109.174103}
}

@article{Arora2020,
  title = {Dislocation Pattern Formation in Finite Deformation Crystal Plasticity},
  author = {Arora, Rajat and Acharya, Amit},
  year = {2020},
  month = feb,
  journal = {International Journal of Solids and Structures},
  volume = {184},
  eprint = {1812.00255},
  pages = {114--135},
  publisher = {Elsevier Ltd},
  issn = {00207683},
  doi = {10.1016/j.ijsolstr.2019.02.013},
  urldate = {2021-02-25},
  archiveprefix = {arXiv}
}

@article{Arsenlis1999,
  title = {Crystallographic Aspects of Geometrically-Necessary and Statistically-Stored Dislocation Density},
  author = {Arsenlis, A. and Parks, D.M},
  year = {1999},
  month = mar,
  journal = {Acta Materialia},
  volume = {47},
  number = {5},
  pages = {1597--1611},
  publisher = {Pergamon},
  issn = {13596454},
  doi = {10.1016/S1359-6454(99)00020-8},
  urldate = {2022-06-30}
}

@article{baileyEmergenceWrappedCauchy2021,
  title = {Emergence of the Wrapped {{Cauchy}} Distribution in Mixed Directional Data},
  author = {Bailey, Joseph D. and Codling, Edward A.},
  year = {2021},
  month = jun,
  journal = {AStA Advances in Statistical Analysis},
  volume = {105},
  number = {2},
  pages = {229--246},
  publisher = {{Springer Science and Business Media Deutschland GmbH}},
  issn = {1863818X},
  doi = {10.1007/s10182-020-00380-7}
}

@article{Birdsall1969,
  title = {Clouds-in-Clouds, Clouds-in-Cells Physics for Many-Body Plasma Simulation},
  author = {Birdsall, Charles K. and Fuss, Dieter},
  year = {1969},
  month = apr,
  journal = {Journal of Computational Physics},
  volume = {3},
  number = {4},
  pages = {494--511},
  issn = {00219991},
  doi = {10.1016/0021-9991(69)90058-8},
  urldate = {2020-01-20}
}

@article{borgiSimulationsDislocationContrast2024,
  title = {Simulations of Dislocation Contrast in Dark-Field {{X-ray}} Microscopy},
  author = {Borgi, S. and R{\ae}der, T. M. and Carlsen, M. A. and Detlefs, C. and Winther, G. and Poulsen, H. F.},
  year = {2024},
  month = apr,
  journal = {Journal of Applied Crystallography},
  volume = {57},
  number = {2},
  pages = {358--368},
  publisher = {International Union of Crystallography},
  issn = {1600-5767},
  doi = {10.1107/S1600576724001183},
  urldate = {2025-06-11},
  copyright = {https://creativecommons.org/licenses/by/4.0/},
  langid = {english}
}

@article{Cai2006,
  title = {A Non-Singular Continuum Theory of Dislocations},
  author = {Cai, Wei and Arsenlis, Athanasios and Weinberger, Christopher and Bulatov, Vasily},
  year = {2006},
  month = mar,
  journal = {Journal of the Mechanics and Physics of Solids},
  volume = {54},
  number = {3},
  pages = {561--587},
  publisher = {Pergamon},
  issn = {00225096},
  doi = {10.1016/j.jmps.2005.09.005},
  urldate = {2018-10-22}
}

@incollection{Devincre2011,
  title = {Modeling {{Crystal Plasticity}} with {{Dislocation Dynamics Simulations}}: {{The}} '{{microMegas}}' {{Code}}},
  booktitle = {Mechanics of {{Nano-Objects}}},
  author = {Devincre, Benoit and Madec, R and Monnet, Ghiath and Queyreau, Sylvain and Gatti, Riccardo and Kubin, Ladislas},
  year = {2011},
  pages = {81--99},
  publisher = {Presses des MINES},
  urldate = {2020-04-07}
}

@article{DeWit1973,
  title = {Theory of Disclinations: {{II}}. {{Continuous}} and Discrete Disclinations in Anisotropic Elasticity},
  author = {DeWit, R},
  year = {1973},
  journal = {Journal of Research of the National Bureau of Standards. Section A, Physics and Chemistry},
  volume = {77},
  number = {1},
  pages = {49},
  publisher = {{National Institute of Standards and Technology}}
}

@article{dusanispanovityAvalanches2DDislocation2014,
  title = {Avalanches in {{2D Dislocation Systems}}: {{Plastic Yielding Is Not Depinning}}},
  author = {Dus{\'a}n Isp{\'a}novity, P{\'e}ter and Laurson, Lasse and Zaiser, Michael and Groma, Istv{\'a}n and Zapperi, Stefano and Alava, Mikko J.},
  year = {2014},
  journal = {Physical Review Letters},
  doi = {10.1103/PhysRevLett.112.235501},
  urldate = {2019-06-24}
}

@article{dusanispanovityCriticalityRelaxationDislocation2011,
  title = {Criticality of {{Relaxation}} in {{Dislocation Systems}}},
  author = {Dus{\'a}n Isp{\'a}novity, P{\'e}ter and Groma, Istv{\'a}n and Gy{\"o}rgyi, G{\'e}za and Szab{\'o}, P{\'e}ter and Hoffelner, Wolfgang},
  year = {2011},
  journal = {Physical Review Letters},
  volume = {107},
  pages = {085506},
  doi = {10.1103/PhysRevLett.107.085506},
  urldate = {2019-06-24}
}

@article{El-Azab2000a,
  title = {Statistical Mechanics Treatment of the Evolution of Dislocation Distributions in Single Crystals},
  author = {{El-Azab}, Anter},
  year = {2000},
  journal = {Physical Review B},
  volume = {61},
  number = {18},
  pages = {11956--11966},
  issn = {1550235X},
  doi = {10.1103/PhysRevB.61.11956},
  urldate = {2019-11-11}
}

@incollection{El-Azab2018,
  title = {Continuum {{Dislocation Dynamics}}: {{Classical Theory}} and {{Contemporary Models}}},
  booktitle = {Handbook of {{Materials Modeling}}},
  author = {{El-Azab}, Anter and Po, Giacomo},
  year = {2018},
  pages = {1--25},
  publisher = {Springer International Publishing},
  doi = {10.1007/978-3-319-42913-7_18-1}
}

@article{Groma1998,
  title = {X-Ray Line Broadening Due to an Inhomogeneous Dislocation Distribution},
  author = {Groma, Istv{\'a}n},
  year = {1998},
  journal = {Physical Review B},
  volume = {57},
  number = {13},
  pages = {7535--7542},
  issn = {1550235X},
  doi = {10.1103/PhysRevB.57.7535},
  urldate = {2019-08-08}
}

@article{Groma1999,
  title = {Investigation of Dislocation Pattern Formation in a Two-Dimensional Self-Consistent Field Approximation},
  author = {Groma, Istv{\'a}n and Balogh, P.},
  year = {1999},
  month = oct,
  journal = {Acta Materialia},
  volume = {47},
  number = {13},
  pages = {3647--3654},
  publisher = {Pergamon},
  issn = {13596454},
  doi = {10.1016/S1359-6454(99)00215-3},
  urldate = {2018-10-18}
}

@article{Groma2003,
  title = {Spatial Correlations and Higher-Order Gradient Terms in a Continuum Description of Dislocation Dynamics},
  author = {Groma, Istv{\'a}n and Csikor, F. F. and Zaiser, Michael},
  year = {2003},
  month = mar,
  journal = {Acta Materialia},
  volume = {51},
  number = {5},
  pages = {1271--1281},
  publisher = {Pergamon},
  issn = {13596454},
  doi = {10.1016/S1359-6454(02)00517-7},
  urldate = {2018-10-18}
}

@article{Groma2015,
  title = {Scale-Free Phase Field Theory of Dislocations},
  author = {Groma, Istv{\'a}n and Vandrus, Zolt{\'a}n and Isp{\'a}novity, P{\'e}ter Dus{\'a}n},
  year = {2015},
  month = jan,
  journal = {Physical Review Letters},
  volume = {114},
  number = {1},
  pages = {015503},
  publisher = {American Physical Society},
  issn = {10797114},
  doi = {10.1103/PhysRevLett.114.015503},
  urldate = {2019-03-04}
}

@article{Groma2021,
  title = {Dynamics of Curved Dislocation Ensembles},
  author = {Groma, Istv{\'a}n and Isp{\'a}novity, P{\'e}ter Dus{\'a}n and Hochrainer, Thomas},
  year = {2021},
  month = may,
  journal = {Physical Review B},
  volume = {103},
  number = {17},
  eprint = {2012.12560},
  pages = {174101},
  issn = {2469-9950},
  doi = {10.1103/PhysRevB.103.174101},
  urldate = {2021-02-24},
  archiveprefix = {arXiv}
}

@phdthesis{Hochrainer2007,
  title = {Evolving Systems of Curved Dislocations: Mathematical Foundations of a Statistical Theory},
  author = {Hochrainer, Thomas},
  year = {2007},
  doi = {10.13140/RG.2.1.1630.6407},
  school = {Karlsruhe Institute of Technology}
}

@phdthesis{andersonStatisticalFoundationsLine2023,
  title = {The {{Statistical Foundations}} of {{Line Bundle Continuum Dislocation Dynamics}}},
  author = {Anderson, Joseph Pierre},
  year = 2023,
  journal = {ProQuest Dissertations and Theses},
  address = {United States -- Indiana},
  urldate = {2026-01-06},
  copyright = {Database copyright ProQuest LLC; ProQuest does not claim copyright in the individual underlying works.},
  isbn = {9798380721318},
  langid = {english},
  school = {Purdue University}
}

@article{Hochrainer2010,
  title = {Dislocation Transport and Line Length Increase in Averaged Descriptions of Dislocations},
  author = {Hochrainer, Thomas and Zaiser, Michael and Gumbsch, Peter},
  year = {2010},
  month = oct,
  journal = {AIP Conference Proceedings},
  volume = {1168},
  eprint = {1010.2884},
  pages = {1133--1136},
  urldate = {2021-02-23},
  archiveprefix = {arXiv}
}

@article{Hochrainer2015,
  title = {Multipole Expansion of Continuum Dislocations Dynamics in Terms of Alignment Tensors},
  author = {Hochrainer, Thomas},
  year = {2015},
  journal = {Philosophical Magazine},
  volume = {95},
  number = {12},
  pages = {1321--1367},
  issn = {14786443},
  doi = {10.1080/14786435.2015.1026297},
  urldate = {2018-11-12}
}

@article{Hochrainer2020,
  title = {Is Crystal Plasticity Non-Conservative? {{Lessons}} from Large Deformation Continuum Dislocation Theory},
  author = {Hochrainer, Thomas and Weger, Benedikt},
  year = {2020},
  month = aug,
  journal = {Journal of the Mechanics and Physics of Solids},
  volume = {141},
  pages = {103957},
  publisher = {Elsevier Ltd},
  issn = {00225096},
  doi = {10.1016/j.jmps.2020.103957},
  urldate = {2021-02-25}
}

@article{Hochrainer2022,
  title = {Making Sense of Dislocation Correlations},
  author = {Hochrainer, Thomas and Weger, Benedikt and Gupta, Satyapriya},
  year = {2022},
  month = dec,
  journal = {Materials Theory},
  volume = {6},
  number = {1},
  pages = {9},
  publisher = {SpringerOpen},
  issn = {2509-8012},
  doi = {10.1186/s41313-021-00040-6},
  urldate = {2022-07-01}
}

@article{Ispanovity2017,
  title = {Role of Weakest Links and System-Size Scaling in Multiscale Modeling of Stochastic Plasticity},
  author = {Isp{\'a}novity, P{\'e}ter Dus{\'a}n and T{\"u}zes, D{\'a}niel and Szab{\'o}, P{\'e}ter and Zaiser, Michael and Groma, Istv{\'a}n},
  year = {2017},
  month = feb,
  journal = {Physical Review B},
  volume = {95},
  number = {5},
  eprint = {1604.01645},
  pages = {054108},
  publisher = {American Physical Society},
  issn = {24699969},
  doi = {10.1103/PhysRevB.95.054108},
  urldate = {2021-02-24},
  archiveprefix = {arXiv}
}

@article{Ispanovity2020,
  title = {Emergence and Role of Dipolar Dislocation Patterns in Discrete and Continuum Formulations of Plasticity},
  author = {Isp{\'a}novity, P{\'e}ter Dus{\'a}n and Papanikolaou, Stefanos and Groma, Istv{\'a}n},
  year = {2020},
  journal = {Physical Review B},
  volume = {101},
  number = {2},
  eprint = {1708.03710v1},
  issn = {24699969},
  doi = {10.1103/PhysRevB.101.024105},
  urldate = {2019-06-25},
  archiveprefix = {arXiv},
  isbn = {1708.03710v1}
}

@article{ispanovityEvolutionCorrelationFunctions2008,
  title = {Evolution of the Correlation Functions in Two-Dimensional Dislocation Systems},
  author = {Isp{\'a}novity, P{\'e}ter Dus{\'a}n and Groma, Istv{\'a}n and Gy{\"o}rgyi, G{\'e}za},
  year = {2008},
  journal = {Physical Review B - Condensed Matter and Materials Physics},
  volume = {78},
  number = {2},
  issn = {10980121},
  doi = {10.1103/PhysRevB.78.024119},
  urldate = {2021-10-29}
}

@article{kocksPhysicsPhenomenologyStrain2003,
  title = {Physics and Phenomenology of Strain Hardening: The {{FCC}} Case},
  shorttitle = {Physics and Phenomenology of Strain Hardening},
  author = {Kocks, U. F. and Mecking, H.},
  year = {2003},
  month = jan,
  journal = {Progress in Materials Science},
  volume = {48},
  number = {3},
  pages = {171--273},
  issn = {0079-6425},
  doi = {10.1016/S0079-6425(02)00003-8},
  urldate = {2025-06-04}
}

@article{Kroner2001,
  title = {Benefits and Shortcomings of the Continuous Theory of Dislocations},
  author = {Kr{\"o}ner, E.},
  year = {2001},
  month = feb,
  journal = {International Journal of Solids and Structures},
  volume = {38},
  number = {6-7},
  pages = {1115--1134},
  publisher = {Pergamon},
  issn = {0020-7683},
  doi = {10.1016/S0020-7683(00)00077-9},
  urldate = {2019-06-14}
}

@article{Lin2020,
  title = {Implementation of Annihilation and Junction Reactions in Vector Density-Based Continuum Dislocation Dynamics},
  author = {Lin, Peng and {El-Azab}, Anter},
  year = {2020},
  month = jun,
  journal = {Modelling and Simulation in Materials Science and Engineering},
  volume = {28},
  number = {4},
  eprint = {1910.12766},
  pages = {045003},
  publisher = {IOP Publishing},
  issn = {0965-0393},
  doi = {10.1088/1361-651X/ab7d90},
  archiveprefix = {arXiv}
}

@article{Lin2021,
  title = {On the Computational Solution of Vector-Density Based Continuum Dislocation Dynamics Models: {{A}} Comparison of Two Plastic Distortion and Stress Update Algorithms},
  author = {Lin, Peng and Vivekanandan, Vignesh and Starkey, Kyle and Anglin, Benjamin and Geller, Clint and {El-Azab}, Anter},
  year = {2021},
  month = mar,
  journal = {International Journal of Plasticity},
  volume = {138},
  pages = {102943},
  publisher = {Elsevier BV},
  issn = {07496419},
  doi = {10.1016/j.ijplas.2021.102943},
  urldate = {2021-02-25}
}

@misc{Lin2021a,
  title = {Crystal Plasticity-Inspired Statistical Analysis of Dislocation Substructures Generated by Continuum Dislocation Dynamics},
  author = {Lin, Peng and Vivekanandan, Vignesh and Castelluccio, Gustavo and Anglin, Benjamin and {El-Azab}, Anter},
  year = {2021},
  month = nov,
  eprint = {2111.12875},
  doi = {10.48550/arxiv.2111.12875},
  urldate = {2022-03-18},
  archiveprefix = {arXiv}
}

@inproceedings{madecNatureAttractiveDislocation2002,
  title = {On the Nature of Attractive Dislocation Crossed States},
  booktitle = {Computational {{Materials Science}}},
  author = {Madec, R. and Devincre, Benoit and Kubin, Ladislas},
  year = {2002},
  month = apr,
  volume = {23},
  pages = {219--224},
  publisher = {Elsevier},
  issn = {09270256},
  doi = {10.1016/S0927-0256(01)00215-4},
  urldate = {2021-02-26}
}

@article{mardiaStatisticsDirectionalData1975,
  title = {Statistics of {{Directional Data}}},
  author = {Mardia, K. V.},
  year = {1975},
  journal = {Journal of the Royal Statistical Society. Series B (Methodological)},
  volume = {37},
  number = {3},
  eprint = {2984782},
  eprinttype = {jstor},
  pages = {349--393},
  publisher = {[Royal Statistical Society, Oxford University Press]},
  issn = {0035-9246},
  urldate = {2025-07-10}
}

@article{Monavari2016,
  title = {Continuum Representation of Systems of Dislocation Lines: {{A}} General Method for Deriving Closed-Form Evolution Equations},
  author = {Monavari, Mehran and Sandfeld, Stefan and Zaiser, Michael},
  year = {2016},
  month = oct,
  journal = {Journal of the Mechanics and Physics of Solids},
  volume = {95},
  eprint = {1509.05617v5},
  pages = {575--601},
  publisher = {Pergamon},
  issn = {00225096},
  doi = {10.1016/j.jmps.2016.05.009},
  urldate = {2019-06-13},
  archiveprefix = {arXiv}
}

@article{Monavari2018,
  title = {Annihilation and Sources in Continuum Dislocation Dynamics},
  author = {Monavari, Mehran and Zaiser, Michael},
  year = {2018},
  month = dec,
  journal = {Materials Theory},
  volume = {2},
  number = {1},
  eprint = {1709.03694},
  pages = {3},
  publisher = {Springer International Publishing},
  issn = {2509-8012},
  doi = {10.1186/s41313-018-0010-z},
  urldate = {2019-02-15},
  archiveprefix = {arXiv}
}

@article{muraContinuousDistributionMoving1963,
  title = {Continuous Distribution of Moving Dislocations},
  author = {Mura, T.},
  year = {1963},
  journal = {Philosophical Magazine},
  volume = {8},
  number = {89},
  pages = {843--857},
  publisher = {Taylor \& Francis Group},
  issn = {00318086},
  doi = {10.1080/14786436308213841},
  urldate = {2021-02-25}
}

@article{Nye1953,
  title = {Some Geometrical Relations in Dislocated Crystals},
  author = {Nye, J. F.},
  year = {1953},
  month = mar,
  journal = {Acta Metallurgica},
  volume = {1},
  number = {2},
  pages = {153--162},
  publisher = {Pergamon},
  issn = {00016160},
  doi = {10.1016/0001-6160(53)90054-6},
  urldate = {2021-02-25}
}

@dataset{dataset,
author = {Joseph P. Anderson},
title = {The line bundle regime and the scale-dependence of continuum dislocation dynamics},
year = 2025,
doi = {10.4231/JW3D-S693} }

@article{Po2018,
  title = {A Non-Singular Theory of Dislocations in Anisotropic Crystals},
  author = {Po, Giacomo and Lazar, Markus and Admal, Nikhil Chandra and Ghoniem, Nasr},
  year = {2018},
  month = apr,
  journal = {International Journal of Plasticity},
  volume = {103},
  eprint = {1706.00828},
  pages = {1--22},
  publisher = {Pergamon},
  issn = {0749-6419},
  doi = {10.1016/J.IJPLAS.2017.10.003},
  urldate = {2022-03-19},
  archiveprefix = {arXiv}
}

@article{poulsenGeometricalopticsFormalismModel2021,
  title = {Geometrical-Optics Formalism to Model Contrast in Dark-Field {{X-ray}} Microscopy},
  author = {Poulsen, H. F. and {Dresselhaus-Marais}, L. E. and Carlsen, M. A. and Detlefs, C. and Winther, G.},
  year = {2021},
  month = dec,
  journal = {Journal of Applied Crystallography},
  volume = {54},
  number = {6},
  pages = {1555--1571},
  publisher = {International Union of Crystallography},
  issn = {1600-5767},
  doi = {10.1107/S1600576721007287},
  urldate = {2024-11-27},
  langid = {english}
}

@phdthesis{Sandfeld2010,
  title = {The {{Evolution}} of {{Dislocation Density}} in a {{Higher-order Continuum Theory}} of {{Dislocation Plasticity}}},
  author = {Sandfeld, Stefan},
  year = {2010},
  school = {University of Edinburgh}
}

@article{Sandfeld2019,
  title = {Orientation-Dependent {{Pattern Formation}} in a 1.{{5D Continuum Model}} of {{Curved Dislocations}}},
  author = {Sandfeld, Stefan and Verbeke, Vanessa and Devincre, Benoit},
  year = {2019},
  journal = {Mater. Res. Soc. Symp. Proc},
  volume = {1},
  pages = {55},
  doi = {10.1557/opl.2015.200},
  urldate = {2019-06-13}
}

@article{SandfeldPo2015,
  title = {Microstructural Comparison of the Kinematics of Discrete and Continuum Dislocations Models},
  author = {Sandfeld, Stefan and Po, Giacomo},
  year = {2015},
  month = dec,
  journal = {Modelling and Simulation in Materials Science and Engineering},
  volume = {23},
  number = {8},
  pages = {085003},
  issn = {0965-0393},
  doi = {10.1088/0965-0393/23/8/085003},
  urldate = {2019-08-09}
}

@article{sauzayScalingLawsDislocation2011,
  title = {Scaling Laws for Dislocation Microstructures in Monotonic and Cyclic Deformation of Fcc Metals},
  author = {Sauzay, M. and Kubin, L. P.},
  year = {2011},
  month = aug,
  journal = {Progress in Materials Science},
  series = {Festschrift {{Vaclav Vitek}}},
  volume = {56},
  number = {6},
  pages = {725--784},
  issn = {0079-6425},
  doi = {10.1016/j.pmatsci.2011.01.006},
  urldate = {2025-06-11}
}

@article{Sedlacek2010,
  title = {The Importance of Being Curved: Bowing Dislocations in a Continuum Description},
  author = {Sedl{\'a}{\v c}ek, Radan and Kratochv{\'i}l, Jan and Werner, Ewald},
  year = {2003},
  month = oct,
  journal = {Philosophical Magazine},
  volume = {83},
  number = {31-34},
  pages = {3735--3752},
  issn = {1478-6435},
  doi = {10.1080/14786430310001600213},
  urldate = {2019-06-17}
}

@article{simonsDarkfieldXrayMicroscopy2015,
  title = {Dark-Field {{X-ray}} Microscopy for Multiscale Structural Characterization},
  author = {Simons, H. and King, A. and Ludwig, W. and Detlefs, C. and Pantleon, W. and Schmidt, S. and St{\"o}hr, F. and Snigireva, I. and Snigirev, A. and Poulsen, H. F.},
  year = {2015},
  month = jan,
  journal = {Nature Communications},
  volume = {6},
  number = {1},
  pages = {6098},
  publisher = {Nature Publishing Group},
  issn = {2041-1723},
  doi = {10.1038/ncomms7098},
  urldate = {2025-06-13},
  copyright = {2015 The Author(s)},
  langid = {english}
}

@article{Song2021,
  title = {Data-Mining of Dislocation Microstructures: Concepts for Coarse-Graining of Internal Energies},
  author = {Song, Hengxu and Gunkelmann, Nina and Po, Giacomo and Sandfeld, Stefan},
  year = {2021},
  month = jan,
  journal = {Modelling and Simulation in Materials Science and Engineering},
  eprint = {2012.14815},
  publisher = {IOP Publishing},
  issn = {0965-0393},
  doi = {10.1088/1361-651x/abdc6b},
  urldate = {2021-02-23},
  archiveprefix = {arXiv}
}

@article{Starkey2020,
  title = {Theoretical Development of Continuum Dislocation Dynamics for Finite-Deformation Crystal Plasticity at the Mesoscale},
  author = {Starkey, Kyle and Winther, Grethe and {El-Azab}, Anter},
  year = {2020},
  month = jun,
  journal = {Journal of the Mechanics and Physics of Solids},
  volume = {139},
  pages = {103926},
  publisher = {Elsevier Ltd},
  issn = {00225096},
  doi = {10.1016/j.jmps.2020.103926},
  urldate = {2020-09-17}
}

@article{Starkey2022,
  title = {Development of Mean-Field Continuum Dislocation Kinematics with Junction Reactions Using de {{Rham}} Currents and Graph Theory},
  author = {Starkey, Kyle and Hochrainer, Thomas and {El-Azab}, Anter},
  year = {2022},
  month = jan,
  journal = {Journal of the Mechanics and Physics of Solids},
  volume = {158},
  pages = {104685},
  publisher = {Pergamon},
  issn = {0022-5096},
  doi = {10.1016/J.JMPS.2021.104685},
  urldate = {2022-03-19}
}

@article{starkeyTotalLagrangeImplementation2022,
  title = {Total {{Lagrange}} Implementation of a Finite-Deformation Continuum Dislocation Dynamics Model of Mesoscale Plasticity},
  author = {Starkey, Kyle and {El-Azab}, Anter},
  year = {2022},
  month = aug,
  journal = {International Journal of Plasticity},
  volume = {155},
  pages = {103332},
  publisher = {Pergamon},
  issn = {0749-6419},
  doi = {10.1016/J.IJPLAS.2022.103332},
  urldate = {2023-03-10}
}

@article{Sudmanns2019,
  title = {Dislocation Multiplication by Cross-Slip and Glissile Reaction in a Dislocation Based Continuum Formulation of Crystal Plasticity},
  author = {Sudmanns, Markus and Stricker, Markus and Weygand, Daniel and Hochrainer, Thomas and Schulz, Katrin},
  year = {2019},
  month = nov,
  journal = {Journal of the Mechanics and Physics of Solids},
  volume = {132},
  pages = {103695},
  publisher = {Elsevier Ltd},
  issn = {00225096},
  doi = {10.1016/j.jmps.2019.103695},
  urldate = {2021-02-23}
}

@article{Sudmanns2020,
  title = {Data-Driven Exploration and Continuum Modeling of Dislocation Networks},
  author = {Sudmanns, Markus and Bach, Jakob and Weygand, Daniel and Schulz, Katrin},
  year = {2020},
  journal = {Modelling and Simulation in Materials Science and Engineering},
  volume = {28},
  number = {6},
  publisher = {IOP Publishing},
  issn = {1361651X},
  doi = {10.1088/1361-651X/ab97ef}
}

@article{Vivekanandan2021,
  title = {On the Implementation of Dislocation Reactions in Continuum Dislocation Dynamics Modeling of Mesoscale Plasticity},
  author = {Vivekanandan, Vignesh and Lin, Peng and Winther, Grethe and {El-Azab}, Anter},
  year = {2021},
  month = apr,
  journal = {Journal of the Mechanics and Physics of Solids},
  volume = {149},
  pages = {104327},
  publisher = {Elsevier Ltd},
  issn = {00225096},
  doi = {10.1016/j.jmps.2021.104327},
  urldate = {2021-02-25}
}

@article{vivekanandanDataDrivenApproach2023,
  title = {A Data Driven Approach for Cross-Slip Modelling in Continuum Dislocation Dynamics},
  author = {Vivekanandan, Vignesh and Anglin, Benjamin and {El-Azab}, Anter},
  year = {2023},
  month = may,
  journal = {International Journal of Plasticity},
  volume = {164},
  pages = {103597},
  issn = {0749-6419},
  doi = {10.1016/j.ijplas.2023.103597},
  urldate = {2025-07-10}
}

@article{Weger2019,
  title = {Leaving the {{Slip System}} - {{Cross Slip}} in {{Continuum Dislocation Dynamics}}},
  author = {Weger, Benedikt and Hochrainer, Thomas},
  year = {2019},
  month = nov,
  journal = {PAMM},
  volume = {19},
  number = {1},
  pages = {201900441},
  publisher = {Wiley},
  issn = {1617-7061},
  doi = {10.1002/pamm.201900441},
  urldate = {2021-02-23}
}

@article{wilkensTheoreticalAspectsKinematical1970,
  title = {Theoretical {{Aspects}} of {{Kinematical X-ray Diffraction Profiles}} from {{Crystals Containing Dislocation Distributions}}},
  author = {Wilkens, M.},
  year = {1970},
  journal = {Fundamental Aspects of Dislocation Theory},
  volume = {II},
  pages = {1195--1221}
}

@article{Wu2018,
  title = {Instability of Dislocation Fluxes in a Single Slip: {{Deterministic}} and Stochastic Models of Dislocation Patterning},
  author = {Wu, Ronghai and T{\"u}zes, Daniel and Isp{\'a}novity, P{\'e}ter Dus{\'a}n and Groma, Istv{\'a}n and Hochrainer, Thomas and Zaiser, Michael},
  year = {2018},
  journal = {Physical Review B},
  volume = {98},
  number = {5},
  pages = {54110},
  issn = {24699969},
  doi = {10.1103/PhysRevB.98.054110},
  urldate = {2019-06-24}
}

@article{wuCellStructureFormation2018,
  title = {Cell Structure Formation in a Two-Dimensional Density-Based Dislocation Dynamics Model},
  author = {Wu, Ronghai and Zaiser, Michael},
  year = {2018},
  month = mar,
  journal = {arXiv},
  eprint = {1803.05951},
  publisher = {arXiv},
  urldate = {2021-02-24},
  archiveprefix = {arXiv}
}

@article{wuCyclicloadingMicrostructurepropertyRelations2019,
  title = {Cyclic-Loading Microstructure-Property Relations from a Mesoscale Perspective: {{An}} Example of Single Crystal {{Nickel-based}} Superalloys},
  author = {Wu, Ronghai and Zaiser, Michael},
  year = {2019},
  month = jan,
  journal = {Journal of Alloys and Compounds},
  volume = {770},
  pages = {964--971},
  publisher = {Elsevier Ltd},
  issn = {09258388},
  doi = {10.1016/j.jallcom.2018.08.168},
  urldate = {2021-02-24}
}

@phdthesis{Xia2015,
  title = {Continuum {{Dislocation Dynamics Modeling}} of the {{Deformation}} of {{FCC Single Crystals}}},
  author = {Xia, Shengxu},
  year = {2015},
  school = {Purdue University}
}

@article{Zaiser2001,
  title = {Statistical Dynamics of Dislocation Systems: {{The}} Influence of Dislocation-Dislocation Correlations},
  author = {Zaiser, Michael and Miguel, M. Carmen and Groma, Istv{\'a}n},
  year = {2001},
  month = nov,
  journal = {Physical Review B},
  volume = {64},
  number = {22},
  pages = {2241021--2241029},
  publisher = {American Physical Society},
  issn = {01631829},
  doi = {10.1103/PhysRevB.64.224102},
  urldate = {2018-11-05}
}

@article{Zaiser2015,
  title = {Local Density Approximation for the Energy Functional of Three-Dimensional Dislocation Systems},
  author = {Zaiser, Michael},
  year = {2015},
  month = nov,
  journal = {Physical Review B},
  volume = {92},
  number = {17},
  eprint = {1508.03652v2},
  pages = {174120},
  publisher = {American Physical Society},
  issn = {1550235X},
  doi = {10.1103/PhysRevB.92.174120},
  urldate = {2018-10-30},
  archiveprefix = {arXiv}
}

@article{zhangContinuumDislocationDynamics2025,
  title = {Continuum Dislocation Dynamics as a Phase Field Theory with Conserved Order Parameters: Formulation and Application to Dislocation Patterning},
  shorttitle = {Continuum Dislocation Dynamics as a Phase Field Theory with Conserved Order Parameters},
  author = {Zhang, Yufan and Wu, Ronghai and Zaiser, Michael},
  year = {2025},
  month = mar,
  journal = {Modelling and Simulation in Materials Science and Engineering},
  volume = {33},
  number = {3},
  pages = {035011},
  publisher = {IOP Publishing},
  issn = {0965-0393},
  doi = {10.1088/1361-651X/adc31f},
  urldate = {2025-06-12},
  langid = {english}
}

@article{kroner1981continuum,
  title={Continuum theory of defects},
  author={Kr{\"o}ner, Ekkehart and others},
  journal={Physics of defects},
  volume={35},
  pages={217--315},
  year={1981},
  publisher={North-Holland, Amsterdam}
}
\end{document}